%% file: Multicast_final.tex
\documentclass[journal,twocolumn,10pt]{IEEEtran}
\usepackage{amsthm,amsmath,comment,bbm,amssymb,color,graphicx}
\usepackage{mathtools}
\usepackage{comment}
\usepackage{url}
\usepackage{caption}
\usepackage{subcaption}
\usepackage{epstopdf}
\usepackage{longtable}
\usepackage{cite}
\usepackage{tikz}
\usepackage{multirow}
\usepackage{multicol}
\usepackage{float}
\usepackage{tablefootnote}
\include{notation_new}

\begin{document}
\title{Rate-Splitting Multiple Access for Overloaded Multi-group Multicast: A First Experimental Study}
\author{Xinze Lyu,~\IEEEmembership{Student Member, IEEE}, Sundar Aditya,~\IEEEmembership{Member, IEEE} and Bruno Clerckx,~\IEEEmembership{Fellow, IEEE}
\thanks{This work was funded by UKRI Impact Acceleration Account (IAA) grant EP/X52556X/1, and UKRI grants EP/X040569/1, EP/Y037197/1, EP/X04047X/1, and EP/Y037243/1.}
\thanks{X.~Lyu, S.~Aditya and B.~Clerckx are with the Dept.~of Electrical and Electronic Engg., Imperial College London, London SW7 2AZ, U.K. (e-mail: \{x.lyu21, s.aditya, b.clerckx\}@imperial.ac.uk).}
}
\maketitle

\begin{abstract}
Multi-group multicast (MGM) is an increasingly important form of multi-user wireless communications with several potential applications, such as video streaming, federated learning, safety-critical vehicular communications, etc. Rate-Splitting Multiple Access (RSMA) is a powerful interference management technique that can, in principle, achieve higher data rates and greater fairness for all types of multi-user wireless communications, including MGM. This paper presents the first-ever experimental evaluation of RSMA-based MGM, as well as the first-ever three-way comparison of RSMA-based, Space Divison Multiple Access (SDMA)-based and Non-Orthogonal Multiple Access (NOMA)-based MGM. Using a measurement setup involving a two-antenna transmitter and two groups of two single-antenna users per group, we consider the problem of realizing throughput (max-min) fairness across groups for each of three multiple access schemes, over nine experimental cases in a line-of-sight environment capturing varying levels of pathloss difference and channel correlation across the groups. Over these cases, we observe that RSMA-based MGM achieves fairness at a higher throughput for each group than SDMA- and NOMA-based MGM. These findings validate RSMA-based MGM's promised gains from the theoretical literature.
\end{abstract}


\section{Introduction}
Multi-group multicast (MGM) is a special case of multi-user wireless communications, wherein a multi-antenna transmitter (TX) jointly communicates with several \emph{groups} of users simultaneously. The $g$-th group (integer $g$) has $N_g$ users, where $N_g \geq 1$ in general, and each user within this group desires the same message from the TX. The special case of a single group corresponds to multicast/broadcast communications, whereas $N_g = 1$ (for each $g$) corresponds to unicast communications.

MGM is a key physical (PHY) layer enabler of several \emph{group} applications, such as video streaming, push-to-talk group radio communications, location-based services, safety-critical vehicular communications, etc \cite{nokia_MGM_2024}. The importance of these applications in future wireless networks is driving ongoing standardization efforts on how best to support multicast/broadcast service (MBS) in future wireless standards, most notably in 3GPP (Third Generation Partnership Project) \cite{3gpp_5g_mbs_rel17}.

Analogous to unicast, MGM can be realized through linear precoding (beamforming) at the TX, but with two key differences that need to be factored into the precoder design:
\begin{itemize}
    \item MGM involves the additional constraint that each user within a group must be able to decode its desired message. In general, a wide range of \emph{spatial geometries} may be possible in terms of group composition, resulting in a great variety of channels experienced by users in a group. However, a special case of considerable practical interest is where users within a group are closely situated and experience \emph{similar} channels -- spanning a spectrum from similar channel strengths at one end, all the way to high spatial correlation/alignment at the other. This could arise in applications like gaming, video conferencing, vehicular swarms, etc. Hence, we focus on this special case in this paper.

    \item Additionally, in many MGM applications such as satellite communications \cite{Multicast_survey_satellite} and Internet-of-Things (IoT) \cite{Multicast_survey_5g}, the number of users {typically exceeds the number of antennas at the TX, which is often referred to as an \emph{overloaded} scenario. Interference management is particularly challenging in overloaded scenarios, since the interference power at all users cannot be suppressed to arbitrarily low levels through linear precoding\footnote{For instance, with $N_T$ antennas at the TX and zero-forcing precoding, the interference can be nulled at no more than $N_T - 1$ users.} \cite{RSMAOverloaded1}. Hence, we further focus on the overloaded scenario in this paper.}
    
\end{itemize}
The above factors have given rise to the following precoder design approaches in the literature, motivated by well-known analogues for unicast:
\begin{itemize}
    \item[a)] \emph{SDMA-based MGM}: Most of the existing literature \cite{karipidis_2008_cochannelMGM, TsungHui_MGM_2008, Christopoulos_2014_TSP,Dong_2020_TSP, karipidis_2007_ULA_RxDirection, Wang_2018_access, bornhorst_TSP_2012_relayMGM, Christopoulos_2015_TWC, Zhu_JSAC_2018, Silva_2009_TVT} on MGM focuses on this approach, which is an extension of Space Division Multiple Access (SDMA) for unicast. Thus, each group is treated as a super-user, resulting in one precoder per group, and the precoders are designed to suppress the power of the \emph{inter-group} interference so that it can be treated as noise. While there are several well-motivated objectives for precoder design in MGM, arguably the most popular in the literature is (max-min) rate fairness across the different groups\footnote{Other objectives for MGM precoder design include (i) minimizing transmit power subject to minimum rate constraints at each group \cite{karipidis_2007_ULA_RxDirection, Wang_2018_access, bornhorst_TSP_2012_relayMGM}, (ii) maximizing sum rate under a per antenna power constraint \cite{Christopoulos_2015_TWC}, and (ii) extending zero-forcing and MMSE precoding for MGM \cite{Silva_2009_TVT}.}. For SDMA-based MGM, this problem has been investigated analytically in many settings, namely \cite{karipidis_2008_cochannelMGM,TsungHui_MGM_2008} (with a sum power constraint), \cite{Christopoulos_2014_TSP} (with a per antenna power constraint), \cite{Dong_2020_TSP} (with low-complexity precoder design for large-scale transmit antennas) and \cite{karipidis_2007_ULA_RxDirection} (for line-of-sight groups), \cite{Wang_2018_access} (for satellite communications), \cite{Zhu_JSAC_2018} (for cooperative integrated terrestrial-satellite communications). However, to the best of our knowledge, an experimental evaluation of SDMA-based MGM does not exist in the literature. The existing experimental studies have focused on implementing MGM at higher layers \cite{mgm_lte_implementation, mgm_80211}, instead of the PHY layer. 
    
    \item[b)] \emph{NOMA-based MGM}: This approach is an extension of power domain Non-Orthogonal Multiple Access (NOMA), and is guided by the principle that decoding the interference rather than treating it as noise is more effective for overloaded scenarios, particularly for unicast \cite{ZhiguoDing_TSP_NOMA_MMA}. Like above, this approach also yields one precoder per group, but for the two-group case, one of the groups employs successive interference cancellation (SIC) to decode and subtract the interference, just like the \emph{stronger user} in unicast NOMA. Precoder optimization\footnote{Note: NOMA-based MGM has also been investigated for single-antenna systems \cite{Wang_ICCC2018_v2x} and multi-antenna systems without precoder optimization
    \cite{Ivari_ICC2020_UCNOMA, Ihsan_NOMAV2X_2021}.} for NOMA-based MGM was investigated in \cite{Choi_NOMA_Multicast_2015, Yi_WCNC2017_Multicast_geocast} (objective: minimizing transmit power subject to minimum rate constraints at each group). However, even for the overloaded scenario, NOMA-based MGM exhibited significant gains over SDMA-based MGM only when there is considerable disparity in channel strength across the groups \cite{Yi_WCNC2017_Multicast_geocast}. Moreover, for groups of closely-spaced users with large angular separation between groups, it is plausible that SDMA-based MGM may still be effective in suppressing inter-group interference by exploiting the spatial domain, even in overloaded scenarios. The potential underperformance of NOMA-based MGM relative to SDMA-based MGM in such scenarios, despite a higher receiver complexity, is due to the former's inability to exploit the multi-antenna multiplexing gain \cite{NOMAineffcient,RSMAOverloaded1}. For a clearer picture of the circumstances under which NOMA-based MGM is superior to SDMA-based MGM (in terms of realizing max-min rate fairness across groups, say), an experimental study of the two schemes over a variety of group spatial geometries is needed, something that is missing in the literature.

        
    \item[c)] \emph{RSMA-based MGM}: {Rooted in the rate-splitting concept originally proposed for the two user single-antenna interference channel \cite{HanKobayashi}, Rate-Splitting Multiple Access (RSMA)  has emerged as a powerful multiple access, interference management, and multi-user strategy for modern multi-antenna communication systems (see \cite{mao2022fundmental}, \cite{RSMA_JSAC_Primer} and the references therein). Inspired by its superiority over both SDMA and NOMA for unicast \cite{Mao2018, RSMAUnifying, lyu2023prototype}, RSMA-based MGM bridges a) and b) above to address their limitations}. Specifically, each group \emph{partially} decodes the inter-group interference and \emph{partially} treats it as noise. All groups have the same signal processing complexity for retrieving their desired messages (one stage of SIC, regardless of the number of groups). Unlike a) and b), this approach yields one more precoder than the number of groups (details in Section~\ref{sec: system model}). Precoder optimization for overloaded RSMA-based MGM\footnote{Precoder optimization for overloaded RSMA \emph{unicast} was investigated in \cite{RSMA_overloaded_Onur, RSMAOverloaded2, junliang_RSMAovl_JWCML}.} was investigated in \cite{RSMAOverloaded1,Yalcin_2020_JVT,Yin_2021_TCOM} (objective: max-min rate fairness across groups), where it was demonstrated analytically that RSMA-based MGM achieved a higher minimum rate than the other two approaches. This was validated using link-level simulations in\cite{Hongzhi_JBC}, \cite{hongzhi_RSMAovl_icc} (under perfect CSIT) and \cite{Yin_ICC_2021, Yin_2021_TCOM, Cui_etal_2023_COML} (for satellite communications). However, for a clearer picture of the extent of such gains, an experimental comparison of the three MGM approaches over a variety of group spatial geometries is needed.
\end{itemize}

Indeed, it is surprising that an experimental evaluation of MGM for any of the above multiple access schemes is missing from the literature. In this paper, we address this gap by conducting the first-ever experimental comparison of the fairness performance of RSMA-, SDMA- and NOMA-based MGM. Our contributions are as follows:
\begin{itemize}
    \item For two groups with two users per group, we formulate the practically relevant \emph{MCS-limited} max-min throughput fairness problem for RSMA-based MGM. In the process, we explain how RSMA-based MGM has an advantage over SDMA- and NOMA-based MGM in terms of realizing max-min fairness through a flexible allocation of the common stream (Section \ref{subsec:stage 2}).
    
    \item Using our RSMA prototype (Section~\ref{sec: RSMA SDR prototype}), we empirically solve the above-defined MCS-limited max-min fairness problem through measurements. In our experiments (Section~\ref{sec: measurement results}), we realize nine cases, with the intention of capturing a wide variety in terms of pathloss difference and spatial correlation between the two groups (subject to closely situated users in each group). Over these nine cases, we observe that RSMA-based MGM achieves fairness at a higher minimum throughput than SDMA- and NOMA-based MGM. This is consistent with theoretical predictions \cite{RSMAOverloaded1,Yalcin_2020_JVT,Hongzhi_JBC}.
\end{itemize}

\subsection{Notation}
Column vectors are represented by lowercase bold letters (e.g., $\mathbf{h}$). $|\cdot|$ denotes the magnitude of scalars and cardinality of sets. $\nbbE[\cdot]$, $(\cdot)^H,~ \|\cdot\|, ~\cup$ and $\emptyset$ denote the expectation operator, the Hermitian operator, the Euclidean norm, set union and the empty set, respectively. $\mathcal{CN}(0,\sigma^2)$ denotes the circularly symmetric complex Gaussian distribution with zero mean and variance $\sigma^2$.

\section{System Model}
\label{sec: system model}
Without loss of generality, we consider MGM involving two groups, with each group comprising two single-antenna users (i.e., $N_1 = N_2 = 2$). Each user belongs to exactly one group, and we assume that the composition of the groups is known beforehand to the TX. For brevity, let $(u, g)$ denote user $u$ in group $g~ (u,g \in \{1,2\})$. All the users within a group desire the same message from the TX, which is communicated over two stages -- in the first stage, the TX acquires channel state information (CSIT), which is used in the second stage to design the RSMA-based MGM precoders and transmit signal. The two stages are described in more detail next.

\subsection{Stage 1: CSIT Acquisition}
\label{subsec:stage 1}
We consider data transmission using OFDM signals over $N_c$ subcarriers. Let $N_t$ denote the number of antenna elements at the TX, and let $\nbh_{u,g}[k] \in \nbbC^{N_t}$ denote the slowly varying, flat fading channel experienced by $(u,g)$ over the $k$-th subcarrier ($k=0,\dots,N_c-1$). Through orthogonal pilots transmitted by the TX, $(u,g)$ obtains an estimate (e.g., least squares) of $\nbh_{u,g}[k]$, denoted by $\hat{\nbh}_{u,g}[k]$. To reduce CSI feedback overhead, $(u, g)$ evaluates the \emph{wideband CSI} by averaging $\hat{\nbh}_{u,g}[k]$ over the subcarriers, i.e.,
\begin{equation}
    \label{eq:CSIT averaging}
    \hat{\nbh}_{u,g} = \frac{1}{N_c}\sum^{N_c-1}_{k=0}\hat{\nbh}_{u,g}[k].
\end{equation}

\subsection{Stage 2: RSMA-based MGM Signal Design}
\label{subsec:stage 2}
 \begin{figure*}
     \centering
     \includegraphics[width=\textwidth]{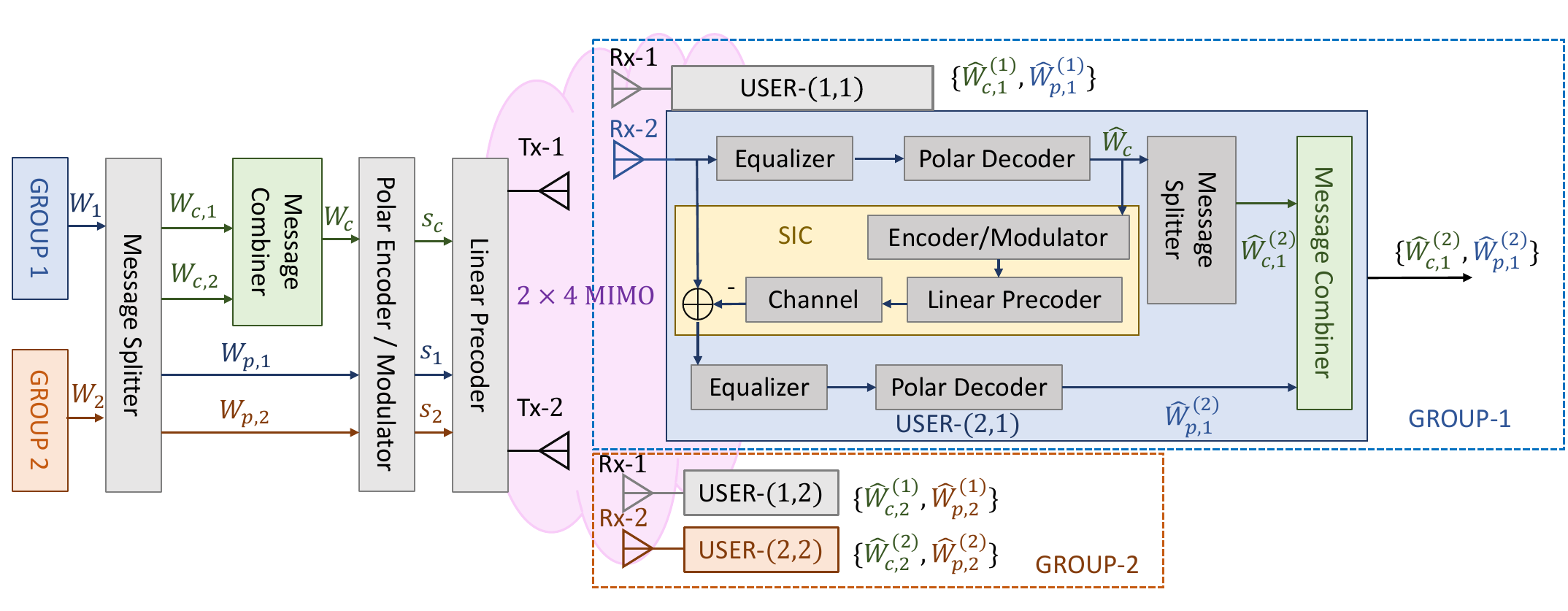}
     \caption{An illustration of RSMA-based MGM.}
     \label{fig:multigroup multicast diagram}
 \end{figure*}
Fig.~\ref{fig:multigroup multicast diagram} depicts RSMA-based MGM in operation \cite{RSMAOverloaded1}. Let $W_1$ and $W_2$ denote the messages meant for groups 1 and 2, respectively. At the TX, each $W_g$ ($g = 1, 2$) is split (by the message splitter block) into common and private portions denoted by $W_{c,g}$ and $W_{p,g}$, respectively (i.e., $W_g = W_{c,g} \cup W_{p,g}$). The two common portions -- $W_{c,1}$ and $W_{c,2}$ -- are then combined (by the message combiner block) into a common message, $W_c$. The three message components -- $W_c$, $W_{p,1}$ and $W_{p,2}$ -- are then individually encoded and modulated to form data streams $s_c[k]$, $s_1[k]$ and $s_2[k]$, respectively, over the subcarriers ($k = 0, \cdots, N_c - 1$). $s_c[\cdot]$ is referred to as the \emph{common stream}, whereas $s_1[\cdot]~(s_2[\cdot])$ is referred to as the \emph{private stream} of group 1 (group 2).

The streams are then linearly precoded, giving rise to the transmit signal $\nbx[k]$ over the $k$-th subcarrier, which can be expressed as follows:
\begin{equation}
    \label{eq:ofdm precoded symbols}
    \mathbf{x}[k]=\mathbf{p}_c s_c[k] + \mathbf{p}_1 s_1[k] + \nbp_2 s_2[k].
\end{equation}
where $\nbp_c$ is referred to as the \emph{common stream precoder} and $\nbp_g$ the \emph{private stream precoder} of group $g$. Let $\nbP=[\nbp_c, ~ \nbp_1,  ~ \nbp_2]$ denote the collection of precoders.

\begin{nrem}
 $\nbP$ is designed as a function of the (imperfect) CSIT acquired in Stage~1 [see (\ref{prob: OP_ideal_RSMA_sr})-(\ref{const: OP_ideal_tx_power})]. However, to avoid messy notation, this functional dependence is not shown explicitly.
\end{nrem}

\begin{nrem}
    Assuming unit symbol power (i.e., $\nbbE[|s_c[k]|^2] = \nbbE[|s_1[k]|^2] = \nbbE[|s_2[k]|^2] = 1$), $\|\nbp_c\|^2$ and $\|\nbp_g\|^2$ represent the power allocated to the common stream and group $g$'s private stream, respectively. In (\ref{eq:ofdm precoded symbols}), the same precoders are applied across all subcarriers, which translates to uniform power allocation over the subcarriers. This suboptimal choice is driven by our use of imperfect CSIT in (\ref{eq:CSIT averaging}) to design $\nbP$. The optimal power allocation -- dictated by waterfilling -- requires knowledge of $\nbh_{u,g}[k]$ at the TX for all $k$.
\end{nrem}

The received signal at $(u,g)$, denoted by $y_{u,g}[k]$, is given by: 
\begin{align}
    \label{eq:received OFDM symbol}
     y_{u,g}[k] &=\mathbf{h}_{u,g}^H[k] \mathbf{x}[k] + n_{u,g}[k] \notag \\
            &=\mathbf{h}_{u,g}^H[k] \mathbf{p}_c s_c[k] + \mathbf{h}_{u,g}^H \mathbf{p}_g s_g[k] + \mathbf{h}_{u,g}^H \mathbf{p}_{g'} s_{g'}[k] \notag \\
            &~ + n_{u,g}[k] 
\end{align}
From the perspective of $(u,g)$, the first and second terms in (\ref{eq:received OFDM symbol}) both contain useful information -- the former represents the common stream component, a part of which is meant solely for group $g$ users, while the latter represents the private stream component meant solely for group $g$ users. In contrast, the third term does not contain useful information, as it captures the private stream component meant for the other group $g' \neq g$. Finally, $n_{u,g}[k]\thicksim\mathcal{CN}(0,\sigma^2)$ denotes the receiver noise at $(u,g)$. 

We assume that precoded pilot\footnote{These pilots play a role similar to the demodulation reference signals (DM-RS) used in LTE and 5G NR.} signals have been interspersed across some subcarriers of $\nbx[k]$ to help $(u,g)$ estimate $\nbh_{u,g}^{H}[k] \nbp_c$ and $\nbh_{u,g}^{H}[k] \nbp_g$ upon receiving $y_{u,g}[k]$, in order to decode the useful information in the first two terms of (\ref{eq:received OFDM symbol}). Using its estimate of $\nbh_{u,g}^{H}[k] \nbp_c$, $(u,g)$ first equalizes (\ref{eq:received OFDM symbol}) to decode $s_c[k]$ and recover $W_c$, by treating the second and third (interference) terms in (\ref{eq:received OFDM symbol}) as noise. The decoded estimate of $W_c$ is simultaneously sent to:
    \begin{itemize}
        \item[(a)] the message splitter block to extract $\hat{W}_{c,g}^{(u)}$, which is the estimate of $W_{c,g}$ at $(u,g)$, and

        \item[(b)] the SIC block, which generates an estimate of $\nbh_{u,g}^H[k] \nbp_c s_c[k]$ and subtracts it from $y_{u,g}[k]$. Using its estimate of $\nbh_{u,g}^{H}[k] \nbp_g$, $(u,g)$ then equalizes the resulting residue to decode $s_g[\cdot]$ and recover $W_{p,g}$, while treating the interference from the third term of (\ref{eq:received OFDM symbol}) as noise. Let $\hat{W}_{p,g}^{(u)}$ denote the decoded estimate of $W_{p,g}$ at $u$. Hence, the estimate of $W_g$ at $(u,g)$, denoted by $\hat{W}_g^{(u)}$, is given by $\hat{W}_g^{(u)} = \hat{W}_{c,g}^{(u)} \cup \hat{W}_{p,g}^{(u)}$.  
    \end{itemize} 

 For the design of $\nbP$, from the perspective of the TX, let $R_{c,g}(\mathbf{P})$ denote the {highest (instantaneous) rate (in bits/s/Hz) supporting} error-free decoding of $s_c[\cdot]$ among the users in group $g$, \emph{under the assumption that $\hat{\nbh}_{u,g}$ in (\ref{eq:CSIT averaging}) constitutes perfect CSIT}. It has the following expression:
\begin{align}
    \label{eq:achievable rate common user i}
         R_{c,g}(\mathbf{P}) = \min_u  \log_2\left( 1+\frac{|\hat{\mathbf{h}}_{u,g}^H \mathbf{p}_c|^2}{\sigma^2 + |\hat{\mathbf{h}}_{u,g}^H \mathbf{p}_1|^2 + |\hat{\mathbf{h}}_{u,g}^H \mathbf{p}_2|^2} \right) \\
         (g \in \{1,2\}), \notag
\end{align}
where the minimum reflects the fact that every user in group $g$ must decode $s_c[\cdot]$.
{
\begin{nrem}[Achievability]
\label{rem:achievability}
 Strictly speaking, $R_{c,g}(\nbP)$ is not \underline{achievable} as it may involve transmitting at a rate that is not supported by the true CSI $\{\nbh_{u,g}[k]: \forall k\}$. The \underline{ergodic rate}, $\nbbE_{\nbh, \hat{\nbh}}[R_{c,g}(\nbP)]$, on the other hand is achievable, where the averaging is over the joint distribution of the true CSI and the CSIT \cite{HamdiMISOImperfectCSIT} (subscripts omitted for simplicity). Since this joint distribution is unknown in practice, the ergodic rate is not a tractable metric for designing $\nbP$. Hence, despite not being technically achievable, $R_{c,g}(\nbP)$ [and similarly, $R_{p,g}(\nbP)$ in (\ref{eq:private achievable rate each user})] is still useful\footnote{The ergodic rate, $\nbbE_{\nbh, \hat{\nbh}}[R_{c,g}(\nbP)]$, can be expressed as $\nbbE_{\hat{\nbh}}[\nbbE_{\nbh|\hat{\nbh}}[R_{c,g}(\nbP)]]$, where the outer expectation is w.r.t the distribution of the CSIT, and the inner expectation is w.r.t the conditional distribution of the true CSI, given the CSIT. As an alternative to the \emph{instantaneous} rate $R_{c,g}(\nbP)$, the term $\nbbE_{\nbh|\hat{\nbh}}[R_{c,g}(\nbP)]$ -- known as the \emph{average rate} (and in general, not achievable) -- can also be used to design $\nbP$ as a function of the CSIT \cite{HamdiMISOImperfectCSIT}. However, like with the ergodic rate, the average rate is also not a tractable metric in practice, since the conditional distribution is unknown and obtaining an empirical estimate at the TX involves considerable overhead.} for designing $\nbP$ [see $OP_{\rm mmf}$ in (\ref{prob: OP_ideal_RSMA_sr})-(\ref{const: OP_ideal_tx_power})]. For such $\nbP$, the empirically achievable rates in our experiments are determined by a suitable choice of modulation and coding scheme (MCS) levels.
\end{nrem}
}

Since $s_c[\cdot]$ contains useful information for users in \emph{both} groups, it follows that both groups must be able to decode $s_c[\cdot]$. Hence, {subject to (\ref{eq:achievable rate common user i}), the highest rate supporting} error-free decoding of $s_c[\cdot]$ by both groups -- denoted by $R_c(\mathbf{P})$ and referred to as the \emph{common stream rate} -- is given by:
\begin{equation}
    \label{eq:achievable rate common all user}
    R_{c}(\mathbf{P}) = \min_ {g \in \{1,2\}} R_{c,g}(\mathbf{P}).
\end{equation}
Assuming perfect SIC (in other words, assuming $R_{c}(\nbP)$ is achievable for all users), let $R_{p,g}(\mathbf{P})$ denote the highest (instantaneous) rate supporting error-free decoding of $s_{g}[\cdot]$ among the users in group $g$, again under the assumption that (\ref{eq:CSIT averaging}) constitutes perfect CSIT. This is referred to as \emph{group $g$'s private stream rate} and has an expression similar to (\ref{eq:achievable rate common user i}), given by: 
\begin{align}
    \label{eq:private achievable rate each user}
    R_{p,g}(\mathbf{P}) = \min_u \log_2 \left( 1+\frac{|\hat{\mathbf{h}}_{u,g}^H \mathbf{p}_g|^2}{\sigma^2 + |\hat{\mathbf{h}}_{u,g}^H \mathbf{p}_{g'}|^2} \right) \\
    ~ (g \in \{1,2\}, g'\neq g). \notag
\end{align}

Subject to (\ref{eq:achievable rate common all user}) and (\ref{eq:private achievable rate each user}), the highest rate supporting error-free recovery of $W_g$ by all users in group $g$  -- denoted by $R_g(\nbP)$ and referred to as the \emph{net group $g$ rate} -- is given by :
\begin{equation}
    \label{eq：maxSE_user1}
    R_g(\mathbf{P})= \frac{|W_{c,g}|}{|W_c|} R_c(\mathbf{P}) + R_{p,g}(\mathbf{P}) \hspace{5mm} (g \in \{1,2\}).
\end{equation}
where the first term is the fraction of the common message ($W_c$) intended for group $g$. 

In general, the sizes of $W_{c,1}$ and $W_{c,2}$ (in bits) in Fig.~\ref{fig:multigroup multicast diagram} need not be equal, which can be used to enable RSMA-based MGM to achieve better fairness. To illustrate this, suppose group 2's private stream rate is smaller than that of group 1 [i.e., $R_{p,2}(\nbP) < R_{p,1}(\nbP)$] because group 2 users have weaker channels than group 1 users. In such a scenario, allocating a larger fraction of the common stream to the weaker group 2 (by having $|W_{c,2}| > |W_{c,1}|$) pushes the net group rates -- $R_1(\nbP)$ and $R_2(\nbP)$ -- towards equality, as captured by the  right-hand side of (\ref{eq：maxSE_user1}). This motivates the following scheme:
\begin{itemize}
    \item[S1)] if $|R_{p,1}(\nbP) - R_{p,2}(\nbP)| \leq R_c (\nbP)$, then the common stream is allocated by adjusting the sizes of $W_{c,1}$ and $W_{c,2}$ such that:
    \begin{align}
     R_{p,1} (\nbP) + \frac{|W_{c,1}|}{|W_c|} R_c (\nbP) = R_{p,2} (\nbP) + \frac{|W_{c,2}|}{|W_c|} R_c (\nbP),  \notag 
    \end{align}
    to ensure that both groups have the same throughput (i.e., achieving max-min fairness);
    
    \item[S2)] otherwise, when $|R_{p,1} (\nbP) - R_{p,2} (\nbP)| > R_c (\nbP)$, the entire common stream is allocated to the group with the smaller private stream rate.
\end{itemize}

The design of $\nbP$ to realize max-min rate fairness across the two groups yields the following optimization problem:
    \begin{align}
        \label{prob: OP_ideal_RSMA_sr}
    OP_{\rm mmf}: \max_{\mathbf{P}} & ~\min \{R_1(\nbP), R_2(\nbP)\} \\
    \label{const: OP_ideal_tx_power}
    \mbox{s.t.} & ~{\mathrm{tr}}(\mathbf{P} \mathbf{P}^H) \leq P_t,
\end{align}
where (\ref{const: OP_ideal_tx_power}) is the TX power constraint. Due to the non-convexity of $OP_{\rm mmf}$, a tractable algorithm that converges to its global optimal solution does not exist. However, a locally optimal solution that is widely recognized as the gold-standard can be obtained using the \emph{WMMSE method} \cite{RSMAOverloaded1}, which takes advantage of the well-known relationship between the achievable rate and the post-equalization symbol mean square error \cite{Christensen_etal_2008}. Let $\mathbf{P}^\mathrm{wm}=[\nbp_c^{\rm wm}, ~\nbp_1^{\rm wm}, ~\nbp_2^{\rm wm}]$ denote the solution to $OP_{\rm mmf}$ obtained by the WMMSE method. However, the resulting max-min fair rate is difficult to realize in practice because, {in addition to Remark~\ref{rem:achievability}}:
\begin{itemize}
    \item [1.] Data transmission is typically restricted to a finite collection of MCS levels, which caps the achievable rate/throughput, and
    
    \item [2.] Error-free decoding of the streams at any user cannot be guaranteed due to a combination of finite block length effects and CSI estimation errors, which further reduces the measured throughput.
\end{itemize}
To account for {imperfect CSIT}, discrete MCS levels and decoding errors, we consider the \emph{MCS-limited} throughput as the metric of interest throughout this paper. An MCS level is defined by a pair $(m, r)$, where positive integer $m$ denotes the bits per constellation symbol (e.g., 2 for QPSK) and $r \in (0,1]$ denotes the code rate. Let $\ncalM=\{(m_c, r_c), (m_1, r_1), (m_2, r_2)\}$ denote the set of  MCS levels chosen for $s_c[\cdot]$, $s_1[\cdot]$ and $s_2[\cdot]$, respectively. Then, the corresponding \emph{MCS-limited} throughputs (measured in bits/s) with WMMSE precoders are given by: 
\begin{align}
    \label{eq:mcstput_sum}
    T_c^{\rm mcs}(\nbP^{\rm wm}, \ncalM) &= B m_c r_c \times \nbbP(\hat{W}_{c} = W_c), \\
    T_{p,1}^{\rm mcs}(\nbP^{\rm wm}, \ncalM)&= B m_1 r_1 \times \nbbP(\hat{W}_{p,1}^{(1)} = \hat{W}_{p,1}^{(2)} = W_{p,1}), \notag \\
    T_{p,2}^{\rm mcs}(\nbP^{\rm wm}, \ncalM)&= B m_2 r_2 \times \nbbP(\hat{W}_{p,2}^{(1)} = \hat{W}_{p,2}^{(2)} = W_{p,2}) \notag.
\end{align}
In the above expressions, $B$ denotes the effective bandwidth\footnote{The effective bandwidth is the portion of the total bandwidth available for data transferring, after accounting for signalling overhead (e.g., the cyclic prefix in OFDM, pilot subcarriers and guard band in IEEE 802.11 standard)}, and the probability terms capture the loss in throughput due to decoding errors -- specifically, the first expression denotes the probability that $W_c$ is correctly decoded by all groups, while the second and third expressions denote the probability that $W_{p,1}$ and $W_{p,2}$ are correctly decoded by groups 1 and 2, respectively.
\begin{nrem}
    The message decoding probabilities in the right-hand side of (\ref{eq:mcstput_sum}) are a function of the chosen MCS level, as well as the SINR at each user, which in turn, depends on $\nbP^{\rm wm}$. This functional dependence is not shown explicitly to avoid messy notation.
\end{nrem}

Similar to (\ref{eq：maxSE_user1}), the \emph{MCS-limited} throughput for group $g$ with WMMSE precoders is given by: 
\begin{equation}
     \label{eq: mcs limited througput group g}
     T_g^{\rm mcs}(\nbP^{\rm wm},\mathcal{M})=\frac{|W_{c,g}|}{|W_c|}T_c^{\rm mcs}(\nbP^{\rm wm}, \ncalM) + T_{p,g}^{\rm mcs}(\nbP^{\rm wm}, \ncalM)
 \end{equation}
Thus, similar to $OP_{\rm mmf}$ in (\ref{prob: OP_ideal_RSMA_sr})-(\ref{const: OP_ideal_tx_power}), the \emph{MCS-limited}  max-min throughput fairness problem for RSMA-based MGM, which is of considerable practical relevance, can be defined as follows:
\begin{align}
    \label{prob: OP_meas}
    OP_{\rm mmf}^{\rm mcs}: 
    \max_{\ncalM} &~ \min \{T^{\rm mcs}_{1}(\nbP^{\rm wm}, \ncalM), T^{\rm mcs}_{2}(\nbP^{\rm wm}, \ncalM)\}\\
    \label{eq:P0_mcs}
    \mbox{s.t.} &~ \ncalM \in \nbbM,
\end{align}
where $\nbbM$ denotes the collection of permissible MCS levels for the three streams, which is typically pre-determined through standards (see Table~\ref{tab: Mcs table} for the $\nbbM$ used in our measurements). The optimal $\ncalM$ for $OP_{\rm mmf}^{\rm mcs}$ is a function of $\nbP^{\rm wm}$ that is difficult to characterize because the message decoding probabilities in (\ref{eq:mcstput_sum}) do not have closed-form expressions in terms of $\nbP^{\rm wm}$ and $\ncalM$. However, these probabilities can be empirically evaluated through measurements. This motivates us to \emph{empirically solve} $OP_{\rm mmf}^{\rm mcs}$ through a brute force search\footnote{A more sophisticated empirical approach involves link adaptation, where the most suitable MCS level is determined by an ARQ-based mechanism \cite{mosquera2023linkadaption}. This is left for future work.} over $\nbbM$, which is the focus of our measurements in Section~\ref{sec: measurement results}. 

We conclude this subsection by noting SDMA- and NOMA-based MGMs are special cases of RSMA-based MGM, corresponding to specific choices of message splitting, as explained in the following remarks.
\begin{nrem}
    \label{rem:sdma is the special case of rs}
SDMA-based MGM is a special case of RSMA-based MGM, where the common stream is turned off -- i.e., $W_g = W_{p,g}, (g=1,2)$ in Fig.~\ref{fig:multigroup multicast diagram}. The corresponding expressions in (\ref{eq:ofdm precoded symbols})-(\ref{eq:P0_mcs}) for SDMA-based MGM can be obtained by setting $\nbp_c = \mathbf{0}$. Essentially, without a common stream, group $g$ users treat the interference from $W_{g'}~(g' \neq g)$ as noise.
\end{nrem}

\begin{nrem}
    \label{rem:noma is the special case of rs}
NOMA-based MGM is a special case of RSMA-based MGM, where the private stream message of one group (group $2$, say) is turned off (i.e., $W_{p,2}=\emptyset$), and no portion of the other message is allocated to the common stream (i.e., $W_1 = W_{p,1}, W_2 = W_{c,2} = W_c$ in Fig.~\ref{fig:multigroup multicast diagram}). The corresponding expressions in (\ref{eq:ofdm precoded symbols})-(\ref{eq:P0_mcs}) can be obtained by setting $\nbp_2 = \mathbf{0}$. For this example, group 1 users fully decode and subtract the interference from $W_2$, whereas group 2 users treat the interference from $W_1$ as noise. In contrast, for RSMA-based MGM, group $g$ users \emph{partially decode} the interference from $W_{g' \neq g}~(W_{c,g'})$ and \emph{partially treat it as noise} ($W_{p,g'}$). 
\end{nrem}

\begin{nrem}
    \label{rem:S2 ture}
    The common stream allocation, as per condition S2 on page 4, is similar to NOMA-based MGM, albeit the weaker group also has a private stream.
\end{nrem}

\section{RSMA Prototype}
\label{sec: RSMA SDR prototype}

\subsection{Hardware setup}
We implement RSMA-based MGM by using our Software Defined Radio (SDR) based RSMA prototype \cite{lyu2023prototype}. The TX and RXs are realized using National Instruments' (NI) USRP 2942 SDR units, which have two antennas/RF chains. Hence, we use three USRP 2942 units, one to realize a two-antenna TX ($N_t = 2$), and the other two to realize four single-antenna users that are assigned to two groups of two users each in our measurements in Section~\ref{sec: measurement results}. In particular, the antennas of group 2 users are connected to their USRP through coaxial cables, which allow them to be moved around to realize different channel environments, as described in Section~\ref{sec: set ups}.
The USRPs share a common timing source (CDA-2990), and are controlled by a workstation running LabVIEW NXG, through which the various blocks in Fig.~\ref{fig:multigroup multicast diagram} are realized. All connections (SDRs to workstation, SDRs to timing sources) are through PCIe cables, facilitated by a PCIe bus (CPS-8910). A list of hardware components is provided in Table \ref{tab: hardware in RSMA}. 

\begin{table}
\centering
\begin{tabular}{|l|p{30mm}|p{40mm}|}
\hline
  & \textbf{Name} & \textbf{Description} \\ 
\hline
 1. & Workstation  & Running LabVIEW NXG     \\ 
 2. & NI USRP-2942 (3 units) & SDRs used to realize TX and users    \\
 3. & NI CPS-8910  & Provides additional PCIe ports\\
 4. & NI CDA-2990 & 8 Channel, 10 MHz clock distribution device \\ 
 5. & 8dBi RP-SMA Male Wifi Antenna&  TX antennas  \\
 6. & RP-SMA Male Wifi Antenna  & Group 1 users' antennas\\
 7. & Mini-Circuits ZAPD-2-272-S+ & Power splitter \\
 8. & TP-Link TL-ANT2405C &  Group 2 users' antennas  \\ \cline{1-3}
\end{tabular}
\caption{List of hardware components.}
\label{tab: hardware in RSMA}
\end{table}

\subsection{RSMA-based MGM Implementation}
\begin{figure*}
    \centering
    \includegraphics[width=0.8\linewidth]{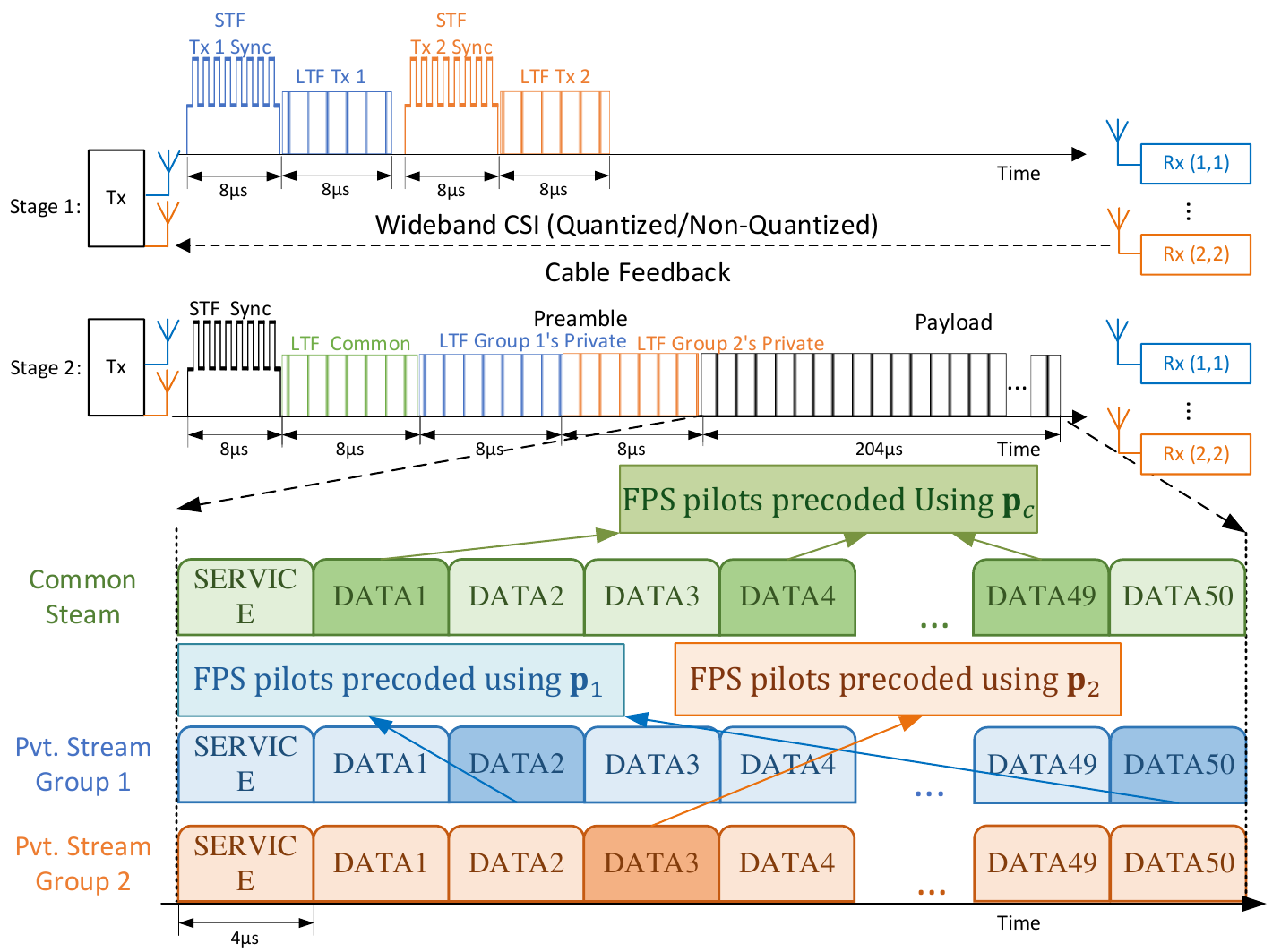}
    \caption{Signal structure within the two-stage transmission protocol used to implement RSMA-based MGM. In conventional (SDMA-based) 802.11g, every DATA (OFDM) symbol contains four \emph{precoded} pilot subcarriers used to correct phase errors in the estimate of a user's precoded channel. This provides protection against decoding errors caused by rotated constellations post equalization, and is known as fine phase shift (FPS) \cite{FinephaseShifting}. However, in RSMA-based MGM, a user is associated with two precoders in general. Hence, we use $\nbp_c$ to precode the FPS pilots in DATA symbols 1, 4, 7 etc., $\nbp_1$ to precode the FPS pilots in DATA symbols 2, 5, 8 etc., and $\nbp_2$ to precode the FPS pilots in DATA symbols 3, 6, 9 etc. This is because the existence of the common/private stream precoders is channel-dependent, and the above scheme ensures that a user can correct phase errors at least once every 3 DATA symbols.}
    \label{fig:ofdm signal stages}
     
\end{figure*}

We adopt several features of the IEEE 802.11g physical layer frames to implement the system model described in Section~\ref{sec: system model}.
\paragraph{Stage 1}
Each TX antenna transmits a pilot signal orthogonally in time comprising a Short Training Field (STF, $8\mu{\rm s}$ in duration) and a Long Training Field (LTF, $8\mu{\rm s}$ in duration), as shown in the top portion of Fig.~\ref{fig:ofdm signal stages}. The STF is used for synchronization and coarse frequency offset estimation, while the LTF is used for least squares based CSI estimation at each user (i.e., to obtain $\hat{\nbh}_{u,g}[k]$ in Section~\ref{subsec:stage 1}).

\paragraph{Stage 2}
The transmitted signal consists of a preamble followed by the data payload, as shown in the bottom portion of Fig.~\ref{fig:ofdm signal stages}.  
    \begin{itemize}
        \item \textbf{Preamble}: The preamble consists of one STF and three LTFs. The function of the STF is the same as in Stage~1, while the LTFs are precoded in order to estimate the \emph{precoded CSI} for equalization at the users. The first LTF is used by each $(u,g)$ to estimate $\nbh_{u,g}^H \nbp_c$ for decoding the common stream. The second LTF is used by each user $u$ in group 1 to estimate $\nbh_{u,1}^H \nbp_1$ to decode its private stream. Similarly, the third LTF is used by each user $u'$ in group 2 to estimate $\nbh_{u',2}^H \nbp_2$ to decode its private stream.
        
        \item \textbf{Data Payload}: For the payload, we consider a total bandwidth of $20{\rm MHz}$ with $N_c = 64$ subcarriers and a cyclic prefix (CP) of 16 samples per OFDM symbol. Aligned with IEEE 802.11 frames, $52$ subcarriers are used for communications while the rest serve as guard bands. Among these $52$ subcarriers, $48$ are used to carry data symbols, with the remaining are pilot subcarriers for FPS, which are used to correct the common phase error across all subcarriers in one OFDM symbol \cite{FinephaseShifting}. This yields an effective bandwidth of:
        \begin{align}
        \label{eq:Beff}
            B = 20{\rm MHz} \times \underbrace{(64/80)}_\text{CP overhead} \times \underbrace{(48/64)}_{\substack{\text{Guard band} \\ \text{overhead}}} = 12{\rm MHz} 
        \end{align}
        The payload consists of three superposed streams (one common, two private), each comprising 50 OFDM symbols.

        \item \textbf{MCS Implementation}: Table~\ref{tab: Mcs table} lists the MCS levels, $\nbbM$, implemented in our prototype. For channel coding, we implement Polar codes augmented with an 8-bit cyclic redundancy check \cite{trifonovPolar, constructionPolar}, along with successive cancellation list decoding \cite{listdecoding}, with a list depth of 2. After the preamble, the first OFDM symbol (labelled SERVICE in the bottom portion of Fig.~\ref{fig:ofdm signal stages}) contains the MCS information of each stream. 
        
        \end{itemize}
        \begin{table}
        \centering
        \begin{tabular}{|c|c|c|c|}
        \hline
        MCS   & Modulation                & Code Rate & Data Rate \\ 
        Index       & ($m$)       & $r$       &  $Bmr$ (Mbps)\\
        \hline
        $0$         & BPSK (1)    & $1/2$         & $6$               \\ \hline
        $1$         & BPSK (1)    & $3/4$         & $9$               \\ \hline
        $2$         & QPSK (2)    & $1/2$         & $12$              \\ \hline
        $3$         & QPSK (2)    & $3/4$         & $18$              \\ \hline
        $4$         & 16QAM (4)   & $1/2$         & $24$              \\ \hline
        $5$         & 16QAM (4)   & $3/4$         & $36$              \\ \hline
        $6$         & 64QAM (6)   & $2/3$         & $48$              \\ \hline
        $7$         & 64QAM (6)   & $3/4$         & $54$              \\ \hline
        $8$         & 256QAM (8)  & $3/4$         & $72$              \\ \hline
        $9$         & 256QAM (8)  & $5/6$         & $80$              \\ \hline
        \end{tabular}
        \caption{MCS levels (largely based on IEEE 802.11g) implemented in our prototype. The data rate in the last column is equal to $B m r$, where $B$ is the effective bandwidth given by (\ref{eq:Beff}).}
         \label{tab: Mcs table}
        \end{table}

    An instance of Stage~1 and Stage~2, as described above and illustrated in Fig.~\ref{fig:ofdm signal stages}, constitutes a single measurement run. To empirically solve $OP_{\rm mmf}^{\rm mcs}$, we conduct $100$ measurement runs. Let $D_{c}$ denote the number of runs in which the common stream is successfully decoded by all users. Similarly, let $D_{g}~(g \in\{1,2\} )$ denote the number of runs in which both users in group $g$ successfully decode their private stream. Replacing the message decoding probabilities in (\ref{eq:mcstput_sum}) with their empirical estimates, the \emph{measured} common and private stream throughputs for RSMA-based MGM is given by: 
        \begin{align}
            \label{eq: throughput calculation in RSMA SDR measurement}
            T^{\rm mcs}_c(\nbP^{\rm wm}, \ncalM) &= \frac{D_c}{100} B m_c r_c, \notag \\
            T^{\rm mcs}_{p,1}(\nbP^{\rm wm}, \ncalM) &= \frac{D_1}{100} B m_1 r_1, \notag \\
            T^{\rm mcs}_{p,2}(\nbP^{\rm wm}, \ncalM) &= \frac{D_2}{100} B m_2 r_2,  
        \end{align}
where $B$ is given by (\ref{eq:Beff}) and the MCS levels are chosen from Table~\ref{tab: Mcs table}.

\begin{table}[h]
 \centering
    \begin{tabular}{|l|c|c|}
    \hline 
      Parameter & Notation  & Value\\
    \hline
       Center frequency  & $f_c$  & $2.484{\rm GHz}$\tablefootnote{This value corresponds to channel no. $14$ in the IEEE 802.11 family of standards for the $2.4{\rm GHz}$ band. We use this channel to avoid ambient WiFi interference, as it is not commercially used in the UK.}\\
       Transmit power & $P_t$ & $23{\rm dBm}$\\
       No. of groups & & 2 \\
       No. of users per group & & 2 \\
    \hline   
       Total bandwidth & & $20{\rm MHz}$ \\
       Subcarriers & Total ($N_c$) & $64$ \\
       & Data & $48$ \\
       & Pilot (FPS) & $4$\\
       & Guard band  & $12$ \\
       CP length & & $16$ \\
       Effective bandwidth & $B$ & $12{\rm MHz}$ \\
    \hline
       OFDM symbols in payload & & $50$ \\
       Experiment runs (per case) & & $100$\\
    \hline
       TX antenna spacing &  & $0.13{\rm m}$ \\
       Fraunhofer distance & & $0.28{\rm m}$\\
       Distance between Group-1 RXs and TX & & 1.00m\\
       Distance between Group-2 RXs and TX  & Case 1-3 & 1.00-1.50m\\
       & Case 4-6 & 2.00-2.30m\\
       & Case 7-9 & 3.20-3.50m\\
    \hline 
    \end{tabular}
    \caption{List of parameters used in our experiments.}
    \label{tab:param_list}
\end{table}
{
\begin{nrem}[Message-splitting]
    To demonstrate the gains of RSMA in overloaded MGM scenarios, it is sufficient to demonstrate that all users can decode and subtract the \textbf{entirety} of the common stream while treating the interference from both the private streams as noise. For this, it is sufficient and much simpler to implement the common and the two private streams as three separate data streams, rather than obtaining them from two data streams via message-splitting. Hence, in our implementation, we adopt this approach and retroactively assign a fraction of the measured common stream throughput to each groups, as per S1 and S2 on page 4 to mimic message-splitting at the TX. Of course, in a real-world implementation, message-splitting would need to take place.
\end{nrem}
}
\section{Results}
\label{sec: measurement results}

\subsection{Measurement setup}
\label{sec: set ups}
\begin{figure}
    \centering
    \begin{subfigure}{0.47\textwidth}
        \centering
        \includegraphics[width = \linewidth]{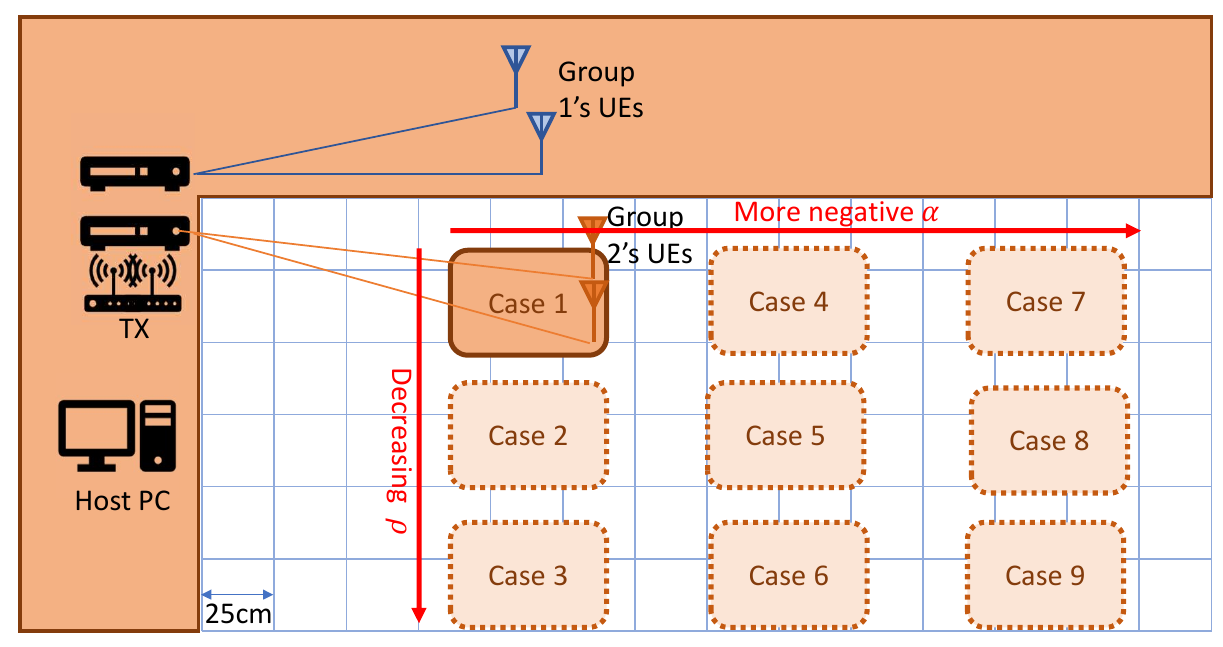}
        \caption{The layout of the measurement environment, along with the TX and user positions for the nine cases. Intuitively, moving in the horizontal direction should increase the inter-group pathloss difference (i.e., $\alpha$ becoming more negative, as per (\ref{eq:pathloss diff alpha})). Likewise, moving in the vertical direction should decrease the spatial correlation (i.e., $\rho$ tending towards 0, as per (\ref{eq:channel correlation})).}
        \label{fig: cases diagram}
    \end{subfigure}
   \begin{subfigure}{0.47\textwidth}
        \centering
        \includegraphics[width = 0.8\linewidth]{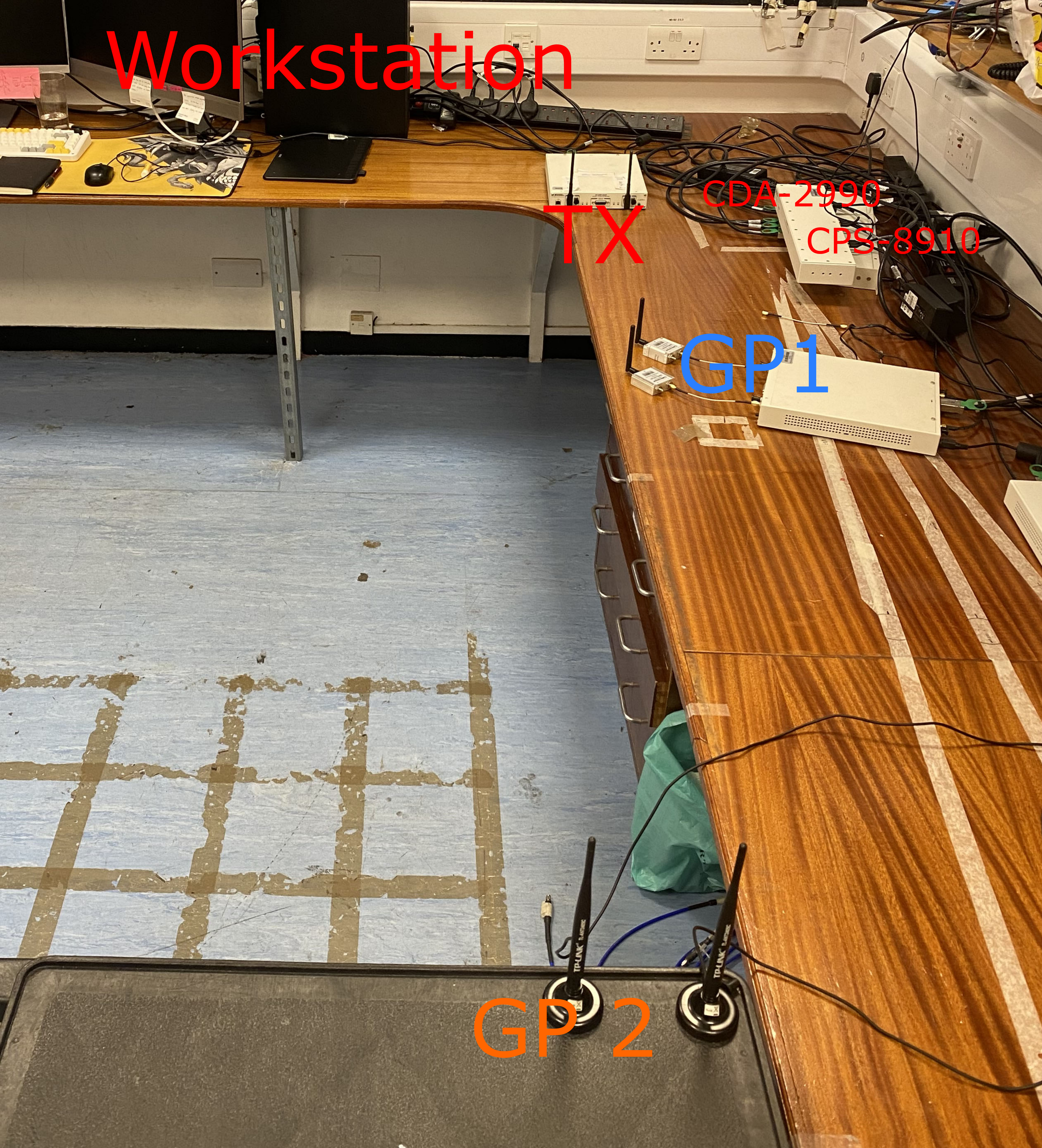}
        \caption{The measurement environment (Case 4 is shown here).}
         \label{fig: cases picture}
    \end{subfigure}
    \caption{Measurement setup}
    \label{fig: measurement cases picture and diagram}    
\end{figure}

In MGM, it is well known that a group's throughput is limited by its \emph{weakest} user. Hence, in the existing literature, there is an assumption of admission control during the formation of the groups, so that users within a group have similar channel strength/link SNR. This motivates us to focus on the scenario where users within a group are closely situated, which could occur in applications like gaming, video conferencing, etc. In such cases, users within a group are likely to experience similar pathloss  and large spatial correlations, whereas there could be considerable pathloss and spatial correlation differences across groups. Furthermore, as the relative performance of SDMA-, NOMA- and RSMA-based MGM is a function of the inter-group interference, it is important to realize a range of environments with varying levels of inter-group interference for a meaningful three-way comparison. This is intrinsically challenging for MGM, as notions of \emph{weak}/\emph{medium}/\emph{strong} interference need to be defined over four pairs of links across the two groups instead of a single pair of links for two-user unicast. Motivated by our unicast measurements \cite[Section IV]{lyu2023prototype}, we rely on geometric notions of channel spatial correlation in line-of-sight environments as a predictor of inter-group interference levels. Thus, similar to \cite{lyu2023prototype}, 
 we consider nine measurement cases, as illustrated in Fig.~\ref{fig: cases diagram},  where moving in the horizontal direction \emph{should} increase the absolute value of inter-group pathloss difference, while moving in the vertical direction is \emph{should} decrease the spatial correlation (and in turn, the interference levels) between the groups.

 To quantify the inter-group pathloss difference between user $u$ in group 1 and user $u'$ in group 2, we define a parameter, $\alpha_{u,u'}$ (in dB scale) as follows:
 \begin{equation}
     \label{eq:pathloss diff alpha}
     \alpha_{u,u'}~ [\mathrm{dB}]=10\log_{10}\frac{\|\hat{\mathbf{h}}_{u',2}\|}{\|\hat{\mathbf{h}}_{u,1}\|},
 \end{equation}
 where $\hat{\mathbf{h}}_{u,g}$ is given by (\ref{eq:CSIT averaging}). {Since group 2 is farther from the TX than group 1, $\alpha_{u,u'}$ is non-positive and thus, a large negative value signifies greater pathloss difference among the two groups}. Likewise, let $\rho_{u, u'}$ denote the spatial correlation between users $u$ in group 1 and $u'$ in group 2, which is defined as follows:
 \begin{equation}
    \label{eq:channel correlation}
    \rho_{u,u'} = \frac{|\hat{\mathbf{h}}_{u,1}^H\hat{\mathbf{h}}_{u',2}|}{||\hat{\mathbf{h}}_{u,1}||\cdot||\hat{\mathbf{h}}_{u',2}||}.
 \end{equation}
Thus, $\rho_{u, u'} = 1$ signifies fully correlated (i.e., highly interfering) channels, while $\rho_{u, u'} = 0$ signifies orthogonal (i.e., zero interference) channels\footnote{ The intra-group spatial correlation can also be obtained from (\ref{eq:channel correlation}), when $u$ and $u'$ belong to the same group. However, we found this to be $\approx 1$ throughout, consistent with closely-spaced (slightly larger than half wavelength) group member. Hence, it is not a useful parameter to explain performance differences across different cases.}. {The expected trends for $\alpha$ and $\rho$ are marked in Fig.~\ref{fig: cases diagram}.}

To realize Fig.~\ref{fig: cases diagram}, group 1 users are fixed on a workbench 1m from the TX, while group 2 users are placed on a trolley (Fig.~\ref{fig: cases picture}) and moved around to realize the nine cases. The distance between the TX and group 2 is provided in Table~\ref{tab:param_list}. All users are in the far field. Table~\ref{tab:channel parameters} captures the spread of the four $\alpha$ and $\rho$ values for the nine cases. While the $\alpha$ spread is fairly narrow and follows the expected trend shown in Fig. 3a, the $\rho$ spread is much larger for several cases because the intuition behind strong/medium/high levels of inter-group spatial correlation in Fig.~\ref{fig: cases diagram} was based on only the (dominant) line-of-sight component, whereas the measured $\rho$ is also sensitive to multipath, which is uncorrelated for each user. This highlights the difficulty of precisely controlling the extent of inter-group interference. Nevertheless, a shift towards 1 in the $\rho$ spread consistent with increasing spatial correlation can be discerned when looking at the low and medium category of cases. No such insight can be drawn from the medium and high categories.

A full list of parameters used in our experiments is provided in Table~\ref{tab:param_list}.
\begin{table}[]
\centering
\begin{tabular}{|c|cp{2cm}||cp{2cm}|}
\hline
\multicolumn{1}{|c|}{}  & \multicolumn{2}{p{30mm}||}{Inter-group Pathloss Difference}             & \multicolumn{2}{p{30mm}|}{ Inter-group Spatial Correlation}                                                                     \\ \cline{2-5}
\multicolumn{1}{|c|}{}  & \multicolumn{1}{c|}{Case} & {$\alpha$ spread [-20,0]} & \multicolumn{1}{c|}{Case} & {$\rho$ spread [0,1]} \\ \hline
\multirow{3}{*}{Low}    & \multicolumn{1}{c|}{1}      &  \begin{tikzpicture}
    \scalebox{2}{
    \draw[-] (0,0) -- (1,0);
    \draw[fill = red, -] (1-0.0195, -0.05) rectangle (1-0.049, 0.05);
    } \end{tikzpicture}   & \multicolumn{1}{c|}{3}  & \begin{tikzpicture}
    \scalebox{2}{
    \draw[-] (0,0) -- (1,0);
    \draw[fill = red, -] (0.18, -0.05) rectangle (0.62, 0.05);
    }
\end{tikzpicture}\\
                  & \multicolumn{1}{c|}{2}      &  \begin{tikzpicture}
    \scalebox{2}{
    \draw[-] (0,0) -- (1,0);
    \draw[fill = red, -] (1-0.059, -0.05) rectangle (1-0.075, 0.05);
    } \end{tikzpicture} & \multicolumn{1}{c|}{6}  & \begin{tikzpicture}
    \scalebox{2}{
    \draw[-] (0,0) -- (1,0);
    \draw[fill = red, -] (0.17, -0.05) rectangle (0.57, 0.05);
    }
\end{tikzpicture}\\
                        & \multicolumn{1}{c|}{3}      & \begin{tikzpicture}
    \scalebox{2}{
    \draw[-] (0,0) -- (1,0);
    \draw[fill = red, -] (1-0.0815, -0.05) rectangle (1-0.0912, 0.05);
    } \end{tikzpicture} & \multicolumn{1}{c|}{9}     & \begin{tikzpicture}
    \scalebox{2}{
    \draw[-] (0,0) -- (1,0);
    \draw[fill = red, -] (0.15, -0.05) rectangle (0.61, 0.05);
    }
\end{tikzpicture}\\ \hline
\multirow{3}{*}{Medium} & \multicolumn{1}{c|}{4}      &  \begin{tikzpicture}
    \scalebox{2}{
    \draw[-] (0,0) -- (1,0);
    \draw[fill = red, -] (1-0.272, -0.05) rectangle (1-0.315, 0.05);
    } \end{tikzpicture}  & \multicolumn{1}{c|}{2}   & \begin{tikzpicture}
    \scalebox{2}{
    \draw[-] (0,0) -- (1,0);
    \draw[fill = red, -] (0.55, -0.05) rectangle (0.76, 0.05);
    }
\end{tikzpicture}   \\
                        & \multicolumn{1}{c|}{5}      & \begin{tikzpicture}
    \scalebox{2}{
    \draw[-] (0,0) -- (1,0);
    \draw[fill = red, -] (1-0.2745, -0.05) rectangle (1-0.335, 0.05);
    } \end{tikzpicture}  & \multicolumn{1}{c|}{5}    & \begin{tikzpicture}
    \scalebox{2}{
    \draw[-] (0,0) -- (1,0);
    \draw[fill = red, -] (0.24, -0.05) rectangle (0.76, 0.05);
    }
\end{tikzpicture} \\
                        & \multicolumn{1}{c|}{6}      & \begin{tikzpicture}
    \scalebox{2}{
    \draw[-] (0,0) -- (1,0);
    \draw[fill = red, -] (1-0.3145, -0.05) rectangle (1-0.3445, 0.05);
    } \end{tikzpicture} & \multicolumn{1}{c|}{8}     & \begin{tikzpicture}
    \scalebox{2}{
    \draw[-] (0,0) -- (1,0);
    \draw[fill = red, -] (0.39, -0.05) rectangle (0.88, 0.05);
    }
\end{tikzpicture}  \\ \hline
\multirow{3}{*}{High}   & \multicolumn{1}{c|}{7}      &  \begin{tikzpicture}
    \scalebox{2}{
    \draw[-] (0,0) -- (1,0);
    \draw[fill = red, -] (1-0.55, -0.05) rectangle (1-0.75, 0.05);
    } \end{tikzpicture}  & \multicolumn{1}{c|}{1}   & \begin{tikzpicture}
    \scalebox{2}{
    \draw[-] (0,0) -- (1,0);
    \draw[fill = red, -] (0.44, -0.05) rectangle (0.99, 0.05);
    }
\end{tikzpicture}    \\
                        & \multicolumn{1}{c|}{8}      & \begin{tikzpicture}
    \scalebox{2}{
    \draw[-] (0,0) -- (1,0);
    \draw[fill = red, -] (1-0.85, -0.05) rectangle (1-0.95, 0.05);
    } \end{tikzpicture} & \multicolumn{1}{c|}{4}     & \begin{tikzpicture}
    \scalebox{2}{
    \draw[-] (0,0) -- (1,0);
    \draw[fill = red, -] (0.43, -0.05) rectangle (0.84, 0.05);
    }
\end{tikzpicture} \\
                        & \multicolumn{1}{c|}{9}      & \begin{tikzpicture}
    \scalebox{2}{
    \draw[-] (0,0) -- (1,0);
    \draw[fill = red, -] (1-0.823, -0.05) rectangle (1-0.95, 0.05);
    } \end{tikzpicture} & \multicolumn{1}{c|}{7}      &\begin{tikzpicture}
    \scalebox{2}{
    \draw[-] (0,0) -- (1,0);
    \draw[fill = red, -] (0.42, -0.05) rectangle (0.93, 0.05);
    }
\end{tikzpicture}\\ \hline
\end{tabular}
\caption{The rectangles in the third and last columns capture the spread of the measured inter-group pathloss difference ($\alpha_{u,u'}$) and inter-group spatial correlation ($\rho_{u,u'}$). The left (right) end of each rectangle corresponds to the smallest (largest) value for each case. As expected, the $\alpha$ spread becomes more negative as the pathloss difference between the groups increases. But, the $\rho$ spread is large for several cases due to sensitivity to multipath. Despite this, a shift towards 1 consistent with increasing spatial correlation can be discerned when looking at the low and medium categories of cases.}
\label{tab:channel parameters}
\end{table}

\subsection{Fairness Comparison}
\label{sec:fairness comparison}
\begin{figure}
    \centering
    \includegraphics[width=0.9\linewidth]{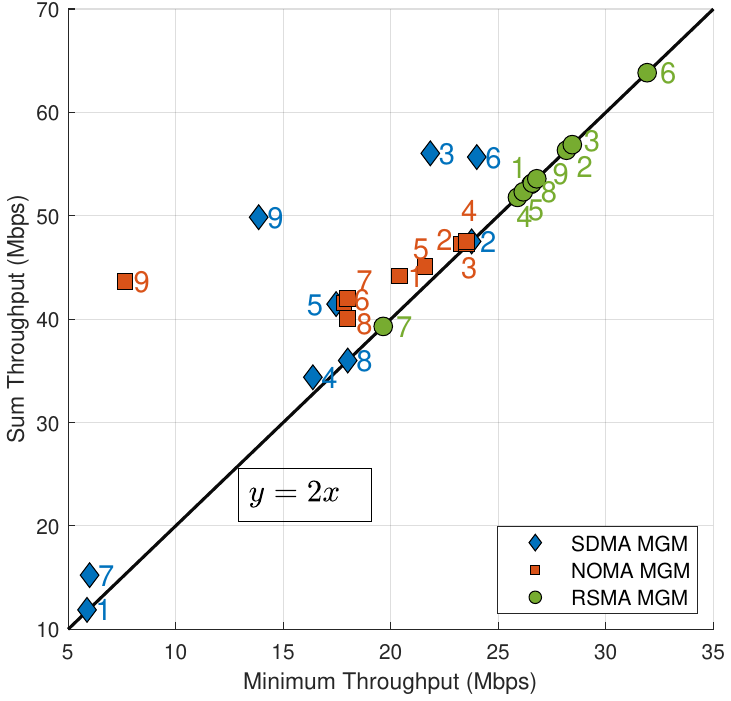}
    \caption{Fairness comparison between SDMA-based, NOMA-based and RSMA-based MGM. The number beside each data point indicates the measurement case. The black line ($y=2x$) corresponds to max-min fairness, and points that are closer (in terms of Euclidean distance) to this line represent fairer outcomes.}
    \label{fig:maxmim scatters}
\end{figure}
\begin{figure}
    \centering
    \begin{subfigure}[b]{0.47\textwidth}
         \centering
    \includegraphics[width=\linewidth]{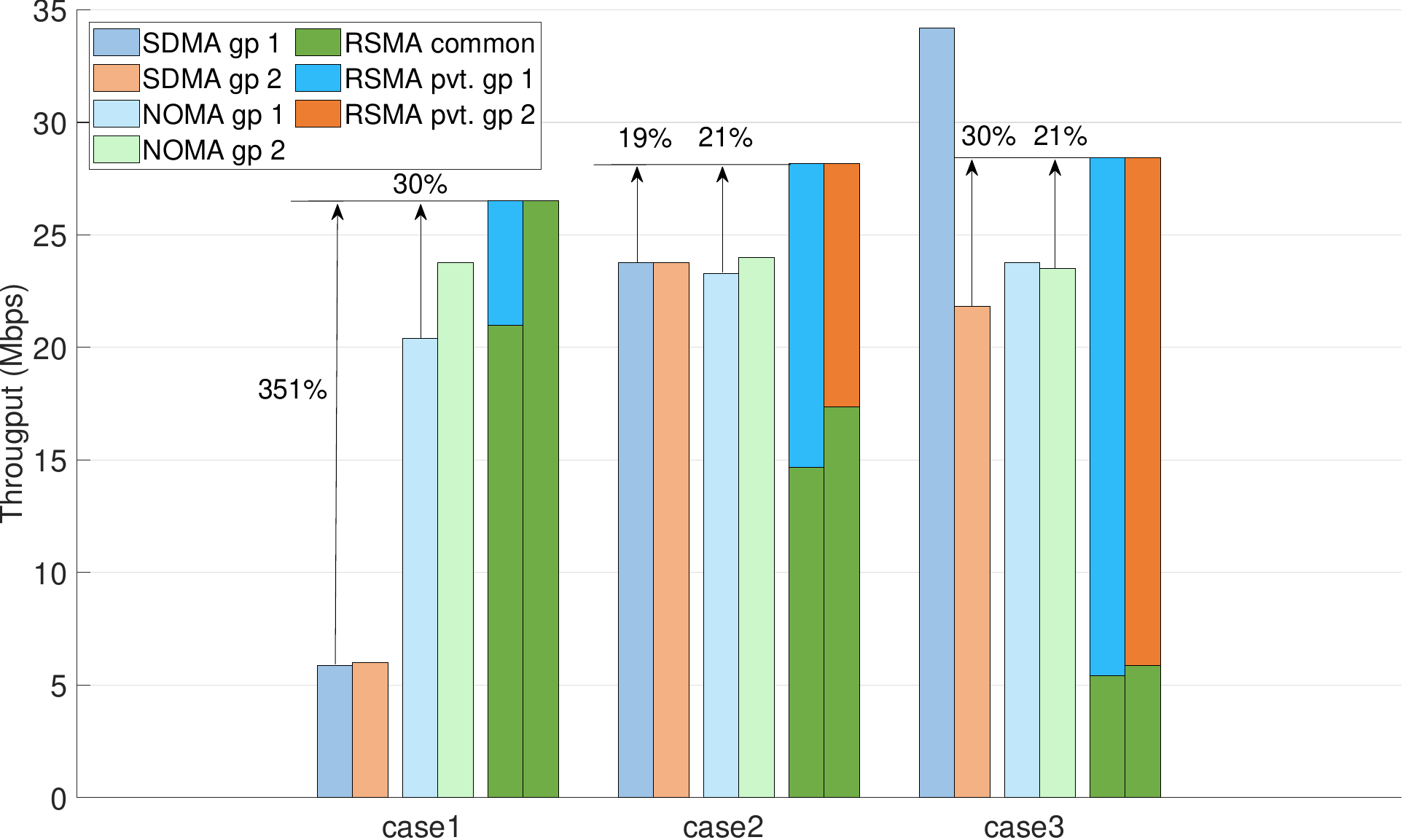}
         \caption{}
         \label{fig:barchart case 13}
     \end{subfigure}
     \hfill
     \begin{subfigure}[b]{0.47\textwidth}
         \centering
        \includegraphics[width=\linewidth]{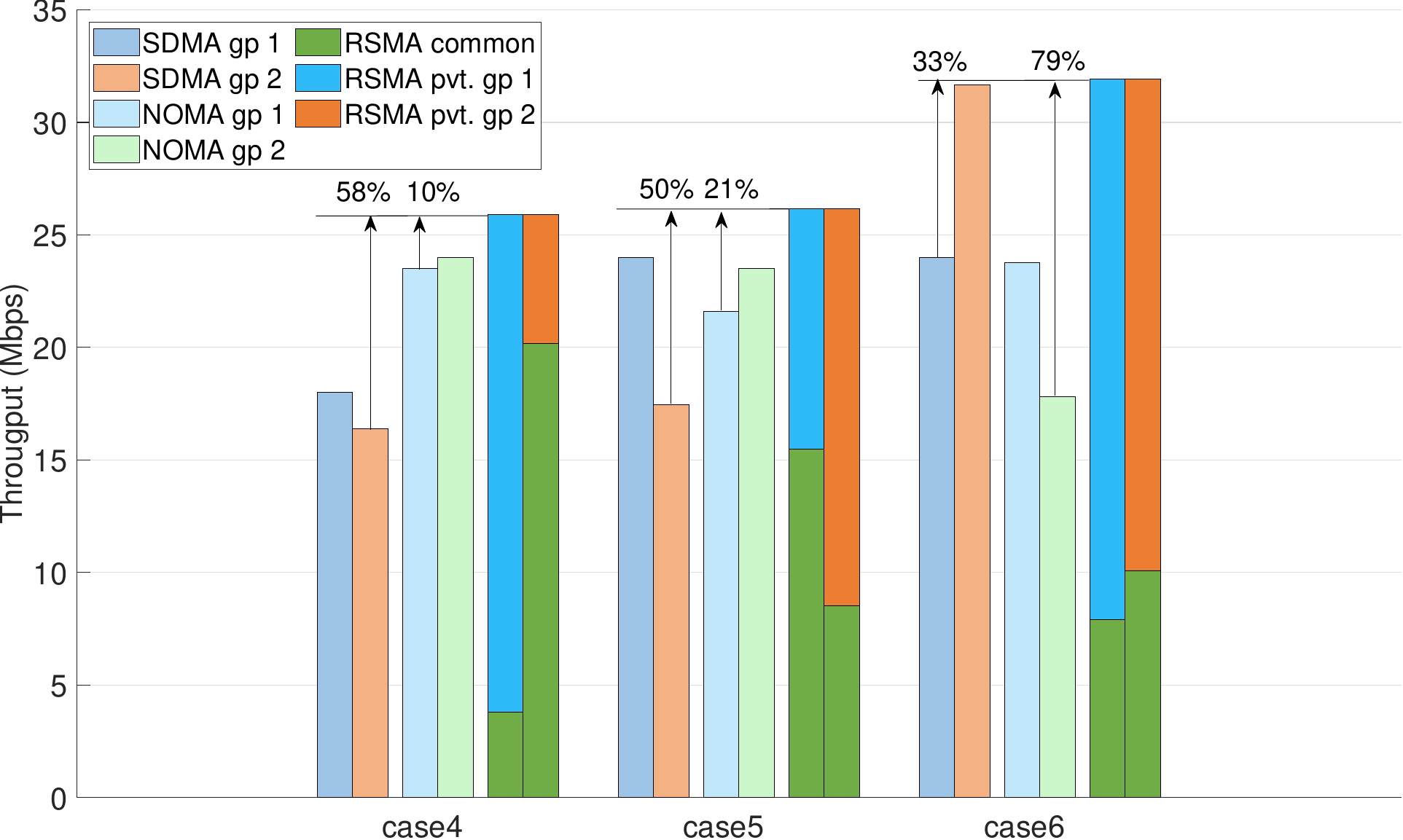}
         \caption{}
         \label{fig:barchart case 46}
     \end{subfigure}
     \hfill
     \begin{subfigure}[b]{0.47\textwidth}
         \centering    \includegraphics[width=\linewidth]{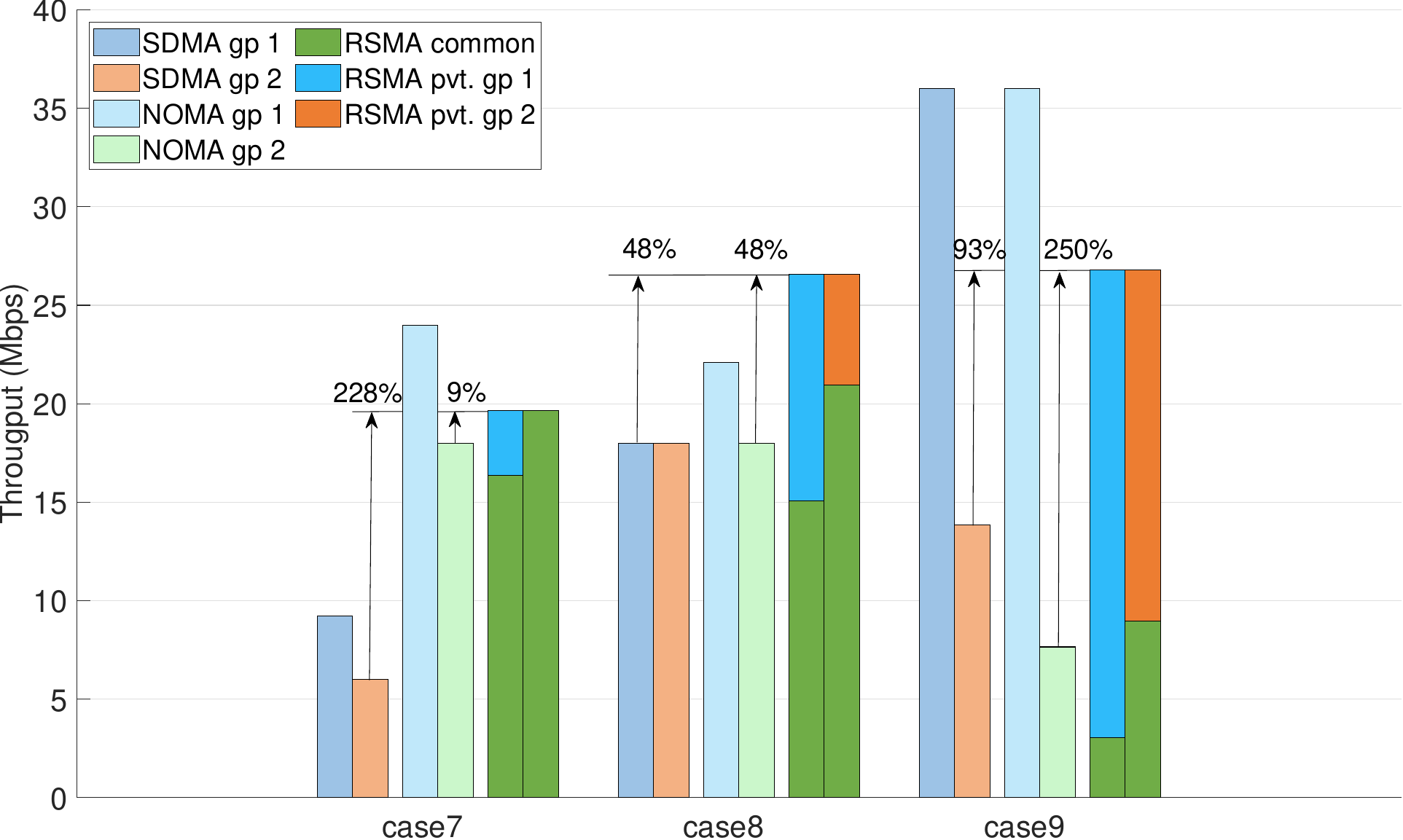}
         \caption{}
         \label{fig:barchart case 79}
     \end{subfigure}
    \caption{The throughput performance for both multicast groups of SDMA-, NOMA- and RSMA-based MGM.}
    \label{fig: throughput bar chart results}
\end{figure}

To compare the fairness performance of RSMA-, NOMA- and SDMA-based MGM, Fig.~\ref{fig:maxmim scatters} plots the sum throughput versus the minimum throughput for each case. The case number is indicated beside each data point. The black $y=2x$ line represents max-min fairness, and the region above is feasible for all three multiple access schemes. Thus, points that are closer (in terms of Euclidean distance) to the black line represent \emph{fairer} outcomes. Furthermore, a point that is either to the north, northeast or east of another point represents a \emph{better} outcome. Based on these insights, we make the following observations:
\begin{itemize}
    \item SDMA: From the spatial geometry of the user locations, cases 1, 4 and 7 have the highest amount of inter-group interference. For these cases, SDMA-based MGM achieves significantly lower minimum and sum throughput than NOMA- and RSMA-based MGM despite achieving near fairness. This is borne by the fact that the SDMA points for these cases lie to the southwest of the corresponding NOMA/RSMA points roughly along the $y=2x$ line. This validates the well-known limitation of treating interference as noise in the high interference regime.

    \item NOMA: With the exception of Case 9, NOMA-based MGM either achieves or comes close to achieving fairness for the other cases. It also achieves better outcomes than SDMA for cases 1, 4, 5, 7 and 8. Interestingly, it achieves strictly \emph{worse} outcomes than SDMA for cases 6 and 9, despite SDMA's limitations on interference suppression in an overloaded scenario. It's worth noting that these cases are associated with the lowest inter-group interference (due to the relatively lower inter-group spatial correlation). This suggests that even for overloaded scenarios where interference cannot be fully suppressed to noise levels, a flexible mix of (partially) decoding the interference and (partially) treating it as noise, in response to the channel conditions is the ideal interference strategy, something RSMA achieves by design.
    
    \item RSMA: In Section~\ref{sec: system model}, when condition S1 (see page 4) is true, the corresponding RSMA-based MGM point lies on the $y=2x$ (i.e., achieves max-min fairness). Otherwise, for condition S2, the RSMA point lies \emph{as close as possible} above the black line. For all cases, RSMA achieves max-min fairness, thereby implying that S1 is satisfied for each case (more on this while discussing Fig. \ref{fig: throughput bar chart results}). Furthermore, compared to SDMA, RSMA achieves strictly better performance for all cases. Likewise, when compared to NOMA, RSMA achieves strictly better performance for all cases bar one -- the exception being case 7, where NOMA achieves a higher sum throughput than RSMA but not fairness (see Section~\ref{subsec:counterintuitive}-c for an explanation). To explain RSMA's superior performance w.r.t SDMA and NOMA, it's important to gain some insight into the relative contributions of the common and private streams to each group's net throughput for each case. We will elaborate on this next.
   
\end{itemize}

Complementing Fig.~\ref{fig:maxmim scatters}, Fig.~\ref{fig: throughput bar chart results} plots the throughput of each group. For RSMA-based MGM, a group's throughput is broken down into common and private stream contributions. {Table~\ref{tab: mcs and bler table} lists the MCS levels and the block error rate (BLER) associated with each bar in Fig.~\ref{fig: throughput bar chart results}}. The arrows indicate the gain in the minimum throughput due to RSMA-based MGM. We make the following observations:
\begin{itemize}
    \item { For SDMA- and NOMA-based MGM, max-min fair precoding allocates more power to the weaker group's precoder in an attempt to equalize the post-precoding SINR at both groups. However, the post-precoding SINR can still be higher at the stronger group, depending on the extent of the inter-group pathloss difference. This SINR mismatch, depending on where it occurs w.r.t an MCS level's SINR boundaries, may permit a higher MCS level to be supported at the stronger group. This explains why fairness is not achieved in some cases for SDMA (3, 5, 7 and 9) and NOMA (6, 7, 8 and 9)\footnote{For the remaining SDMA and NOMA cases, the same MCS level is supported at both groups. The throughput variations for these cases (and any resulting lack of fairness) are due to block errors, which are difficult to control.} -- see Table~\ref{tab: mcs and bler table}}.
    \item For RSMA-based MGM, in each subfigure, the contribution of the common stream to each group's throughput decreases from left to right across the three cases. This pattern was observed for unicast communications, as well \cite{lyu2023prototype}. Essentially, when the inter-group interference is high (as in cases 1, 4 and 7),  most of the power is allocated to the common stream, which, in turn, contributes substantially to the net throughput. In other words, decoding most of the interference is favoured over suppressing it to noise levels. On the other hand, when the inter-group interference is lower (as in cases 3, 6 and 9), suppressing most of the interference to noise levels is preferred and hence, the power allocation to the private streams increases. As a result, the private streams contribute more to the net throughput.
    
    \item For RSMA-based MGM, there is a non-zero common stream contribution to each group's throughput in every case. Thus, S1 (see page 4) is satisfied for each case, and the allocation of the common stream to each group is done to ensure that both groups have the same net throughput. In other words, throughput contributions from both the common and the private streams for each group provides the required flexibility in terms of resource allocation to ensure that fairness is realized for each case.
 
    \item To further highlight the benefits of throughput contributions from both the common and the private streams for each group, consider cases 1 and 7 for RSMA-based MGM. Since the private stream contribution to (the weaker) group 2's throughput is zero, it appears at first glance as though RSMA-based MGM reduces to NOMA-based MGM. But there is a key difference. Unlike NOMA, the common stream can contribute to (the stronger) group 1's throughput as well, to the extent that most of group 1's throughput is due to the common stream. This is intuitive, since under high inter-group interference, most of the power is allocated to the common stream. In contrast, for NOMA-based MGM, group 1 needs to rely entirely on its private stream for its throughput. This forces a suboptimal resource allocation, where power has to be diverted from the common stream to group 1's private stream to achieve fairness. As a result, NOMA-based MGM achieves a lower minimum throughput than RSMA-based MGM for these cases. The flexibility offered by throughput contributions from both the common and the private streams for each group is also why RSMA-based MGM achieves better outcomes than both SDMA- and NOMA-based MGM.
\end{itemize}
In summary, RSMA-based MGM adapts well to variations in pathloss difference and interference between the groups and achieves superior fairness performance (i.e., fairness at a higher minimum throughput) when compared to SDMA- and NOMA-based MGM. We conclude this section by remarking on some seemingly counter-intuitive results in Fig.~\ref{fig: throughput bar chart results} {and Table~\ref{tab: mcs and bler table}}.

\subsection{Counter-intuitive Results}
\label{subsec:counterintuitive}
\begin{table*}[t]
\centering
\begin{tabular}{|c|cc|c c||c c|c c||c c|c c|c c|}
\hline
 & \multicolumn{14}{c|}{BLER (MCS index from Table~\ref{tab: Mcs table})} \\
 \cline{2-15}
Case & \multicolumn{4}{c||}{SDMA}    & \multicolumn{4}{c||}{NOMA}    & \multicolumn{6}{c|}{RSMA}                                   \\ \cline{2-15} 
  & \multicolumn{2}{c|}{Group 1}     & \multicolumn{2}{c||}{Group 2}     & \multicolumn{2}{c|}{Group 1}     & \multicolumn{2}{c||}{Group 2}     & \multicolumn{2}{c|}{Common}  & \multicolumn{2}{c|}{Group 1 pvt.} & \multicolumn{2}{c|}{Group 2 pvt.} \\ \hline
1  & 0.02 & (0) & 0 & (0)    & 0.15 & (4) & 0.01 & (4) & 0.01 & (6) &  0.08 & (0) & -- & --    \\ \hline
2  & 0.01 & (4) & 0.01 & (4) & 0.03 & (4) & 0 & (4) & 0.10 & (5) &  0.25 & (3) & 0.10 & (2) \\ \hline
3 & 0.05 & (5) & 0.09 & (4) & 0.01 & (4) & 0.02 & (4) & 0.06 & (2) & 0.04 & (4) & 0.06 & (4) \\ \hline
4 &  0 & (3)    & 0.09 & (3) & 0.02 & (4) & 0 & (4)    & 0 & (4) & 0.08 & (4) & 0.05 & (0) \\ \hline
5 & 0 & (4) & 0.03 & (3) & 0.11 & (4) & 0.02 & (4) & 0 & (4) & 0.11 & (2) & 0.02 & (3) \\ \hline
6 & 0 & (4) & 0.12 & (5) & 0.01 & (4) & 0.01 & (3) & 0 & (3) & 0 & (4) & 0.09 & (4) \\ \hline
7 & 0.23 & (2) & 0 & (0) & 0 & (4) & 0 & (3) & 0 & (5) & 0.45 & (0) & -- & --    \\ \hline
8 & 0 & (3) & 0 & (3) & 0.08 & (4) & 0 & (3) & 0 & (5) & 0.04 & (2) & 0.06 & (0) \\ \hline
9 & 0 & (5)    & 0.23 & (3) & 0 & (5)    & 0.15 & (1) & 0 & (2)    & 0.01 & (4) & 0.01 & (3) \\ \hline
\end{tabular}
\caption{The MCS Index and the BLER corresponding to each throughput bar in Fig.~\ref{fig: throughput bar chart results}.}
\label{tab: mcs and bler table}
\end{table*}
\paragraph{SDMA, Case 6} Perhaps the most counter-intuitive observation in Fig.~\ref{fig: throughput bar chart results} is the supposedly weaker group 2 having a higher throughput (31.68Mbps) than the stronger group 1 (24Mbps) for SDMA-based MGM in Case 6. {From Table~\ref{tab: mcs and bler table}, the MCS levels and the block error rate (BLER) corresponding to this particular case are}:
\begin{itemize}
    \item Group 1: 16QAM, rate 1/2 (level 4 in Table~\ref{tab: Mcs table}) with 0 BLER
    \item Group 2: 16QAM, rate 3/4 (level 5 in Table~\ref{tab: Mcs table}) with 0.12 BLER
\end{itemize}
It is clear from above that group 2's higher throughput stems from its ability to support a higher MCS level than group 1, albeit it is only one level higher and not error-free. The higher MCS level is due to the fact that more power is allocated to the group 2 precoder to achieve fairness. Ideally, under perfect CSIT, one would expect more power to be allocated to the group 2 precoder in such a way that both groups support the \emph{same} MCS level, thereby achieving the same throughput. However, with imperfect CSIT, which is inherent in our experiments due to the use of wideband CSI in (\ref{eq:CSIT averaging}), we reckon it is not unusual for group 2 to support an MCS level one higher than group 1. Moreover, if we had imposed strict reliability requirements (i.e., zero BLER in our measurements), then both groups would have supported the same MCS level (16QAM, rate 1/2) and achieved the same throughput. 

\paragraph{NOMA, Cases 1, 2, 4 and 5} Again, group 2 has a higher throughput than group 1 for these cases, although the difference is not as stark as above. This is because both groups support the same MCS level (see Table~\ref{tab: mcs and bler table}) which is expected, but the throughput variations are due to block errors that are difficult to control.

\paragraph{NOMA v/s RSMA, Case 7} For NOMA-based MGM, groups 1 and 2 supports MCS levels 4 and 3, respectively (both with zero BLER, see Table~\ref{tab: mcs and bler table}), and the resulting sum throughput is 42Mbps. Due to the high pathloss difference and the high inter-group interference associated with this case, RSMA-based MGM reduces to NOMA-based MGM, as evidenced by the lack of a private stream for group 2 (see Remark~\ref{rem:noma is the special case of rs}). But, unlike NOMA, the RSMA common stream can contribute to the throughput of \emph{both} groups. Hence, nearly all of the transmit power is allocated to the common stream precoder. Consequently, the common stream alone should be able to support an MCS level whose throughput is at least 42Mbps (i.e., the NOMA sum throughput), which would then be equally split between the two groups to achieve fairness. However, from Table~\ref{tab: Mcs table}, we see that there is no MCS level close to 42Mbps. Hence, in Table~\ref{tab: mcs and bler table}, we see that the common stream supports MCS level 5 (36Mbps) which is the nearest level lower than 42Mbps, but not MCS level 6 (48Mbps) which is the nearest level above 42Mbps. This is the reason why the NOMA sum throughput exceeds the RSMA sum throughput. However, even with conservative MCS level 5 for the common stream, RSMA-based MGM achieves a higher minimum throughput than NOMA-based MGM, which is the desired objective.

\section{Conclusion}
\label{sec: summary}
In this paper, we presented the first-ever experimental evaluation of the fairness performance of RSMA-, SDMA- and NOMA-based MGM. We focused our attention on the overloaded scenario with closely located group members, which is relevant for many applications. For two groups with two users per group, we realized nine cases that captured varying levels of interference and pathloss difference between the two groups. Over these nine cases, we observed that RSMA-based MGM achieved superior fairness performance (i.e., fairness at a higher minimum throughput) than SDMA- and NOMA-based MGM. This is consistent with theoretical predictions. The secret behind these gains stems from the fact that each group's throughput could, in general, have contributions from \emph{both} the common and the private streams with the exact amounts of each dictated by the channel conditions.




\bibliographystyle{IEEEtran}

\end{document}

%% file: notation_new.tex
\def\nbh{{\mathbf{h}}}

\def\nbp{{\mathbf{p}}}

\def\nbx{{\mathbf{x}}}

\def\nb0{{\mathbf{0}}}
\def\nb1{{\mathbf{1}}}

\def\nbP{{\mathbf{P}}}

\def\ncalM{{\mathcal{M}}}

\def\nbbC{{\mathbb{C}}}

\def\nbbE{{\mathbb{E}}}

\def\nbbM{{\mathbb{M}}}

\def\nbbP{{\mathbb{P}}}

\newtheorem{nrem}{Remark}

\newtheorem{eg}{Example}










%% file: Multicast_final.bbl
\begin{thebibliography}{10}
\providecommand{\url}[1]{#1}
\csname url@samestyle\endcsname
\providecommand{\newblock}{\relax}
\providecommand{\bibinfo}[2]{#2}
\providecommand{\BIBentrySTDinterwordspacing}{\spaceskip=0pt\relax}
\providecommand{\BIBentryALTinterwordstretchfactor}{4}
\providecommand{\BIBentryALTinterwordspacing}{\spaceskip=\fontdimen2\font plus
\BIBentryALTinterwordstretchfactor\fontdimen3\font minus \fontdimen4\font\relax}
\providecommand{\BIBforeignlanguage}[2]{{%
\expandafter\ifx\csname l@#1\endcsname\relax
\typeout{** WARNING: IEEEtran.bst: No hyphenation pattern has been}%
\typeout{** loaded for the language `#1'. Using the pattern for}%
\typeout{** the default language instead.}%
\else
\language=\csname l@#1\endcsname
\fi
#2}}
\providecommand{\BIBdecl}{\relax}
\BIBdecl

\bibitem{nokia_MGM_2024}
\BIBentryALTinterwordspacing
B.~Elmali, D.~M. Soleymani, and S.~A. Ashraf, ``Empowering public safety services using multicast in {5G}-{A}dvanced,'' Nokia Blog, Jan. 2024. [Online]. Available: \url{https://www.nokia.com/blog/empowering-public-safety-services-using-multicast-in-5g-advanced/}
\BIBentrySTDinterwordspacing

\bibitem{3gpp_5g_mbs_rel17}
{3GPP TS 26.517}, ``{5G} {M}ulticast-{B}roadcast {U}ser {S}ervices; {P}rotocols and {F}ormats (rel. 17),'' Sep. 2023.

\bibitem{Multicast_survey_satellite}
M.~{\'A}. V{\'a}zquez, A.~P{\'e}rez-Neira, D.~Christopoulos, S.~Chatzinotas, B.~Ottersten, P.-D. Arapoglou, A.~Ginesi, and G.~Taricco, ``{P}recoding in {M}ultibeam {S}atellite {C}ommunications: {P}resent and {F}uture {C}hallenges,'' \emph{{IEEE} Trans. Wireless Commun.}, vol.~23, no.~6, pp. 88--95, 2016.

\bibitem{Multicast_survey_5g}
G.~Araniti, M.~Condoluci, P.~Scopelliti, A.~Molinaro, and A.~Iera, ``Multicasting over {E}merging {5G} {N}etworks: {C}hallenges and {P}erspectives,'' \emph{IEEE Network}, vol.~31, no.~2, pp. 80--89, 2017.

\bibitem{RSMAOverloaded1}
H.~Joudeh and B.~Clerckx, ``{R}ate-{S}plitting for {M}ax-{M}in fair multigroup multicast beamforming in overloaded systems,'' \emph{{IEEE} Trans. Wireless Commun.}, vol.~16, no.~11, pp. 7276--7289, 2017.

\bibitem{karipidis_2008_cochannelMGM}
E.~Karipidis, N.~D. Sidiropoulos, and Z.-Q. Luo, ``Quality of service and max-min fair transmit beamforming to multiple cochannel multicast groups,'' \emph{{IEEE} Trans. Signal Process.}, vol.~56, no.~3, pp. 1268--1279, 2008.

\bibitem{TsungHui_MGM_2008}
T.-H. Chang, Z.-Q. Luo, and C.-Y. Chi, ``Approximation bounds for semidefinite relaxation of {M}ax-{M}in-{F}air multicast transmit beamforming problem,'' \emph{{IEEE} Trans. Signal Process.}, vol.~56, no.~8, pp. 3932--3943, 2008.

\bibitem{Christopoulos_2014_TSP}
D.~Christopoulos, S.~Chatzinotas, and B.~Ottersten, ``Weighted fair multicast multigroup beamforming under per-antenna power constraints,'' \emph{{IEEE} Trans. Signal Process.}, vol.~62, no.~19, pp. 5132--5142, 2014.

\bibitem{Dong_2020_TSP}
M.~Dong and Q.~Wang, ``Multi-group multicast beamforming: Optimal structure and efficient algorithms,'' \emph{{IEEE} Trans. Signal Process.}, vol.~68, pp. 3738--3753, 2020.

\bibitem{karipidis_2007_ULA_RxDirection}
E.~Karipidis, N.~D. Sidiropoulos, and Z.-Q. Luo, ``Far-field multicast beamforming for uniform linear antenna arrays,'' \emph{{IEEE} Trans. Signal Process.}, vol.~55, no.~10, pp. 4916--4927, 2007.

\bibitem{Wang_2018_access}
W.~Wang, A.~Liu, Q.~Zhang, L.~You, X.~Gao, and G.~Zheng, ``Robust multigroup multicast transmission for frame-based multi-beam satellite systems,'' \emph{IEEE Access}, vol.~6, pp. 46\,074--46\,083, 2018.

\bibitem{bornhorst_TSP_2012_relayMGM}
N.~Bornhorst, M.~Pesavento, and A.~B. Gershman, ``Distributed beamforming for multi-group multicasting relay networks,'' \emph{IEEE Transactions on Signal Processing}, vol.~60, no.~1, pp. 221--232, 2012.

\bibitem{Christopoulos_2015_TWC}
D.~Christopoulos, S.~Chatzinotas, and B.~Ottersten, ``Multicast multigroup precoding and user scheduling for frame-based satellite communications,'' \emph{{IEEE} Trans. Wireless Commun.}, vol.~14, no.~9, pp. 4695--4707, Sep. 2015.

\bibitem{Zhu_JSAC_2018}
X.~Zhu, C.~Jiang, L.~Yin, L.~Kuang, N.~Ge, and J.~Lu, ``Cooperative multigroup multicast transmission in integrated terrestrial-satellite networks,'' \emph{{IEEE} J. Sel. Areas Commun.}, vol.~36, no.~5, pp. 981--992, 2018.

\bibitem{Silva_2009_TVT}
Y.~C.~B. Silva and A.~Klein, ``Linear transmit beamforming techniques for the multigroup multicast scenario,'' \emph{{IEEE} Trans. Veh. Technol.}, vol.~58, no.~8, pp. 4353--4367, Oct 2009.

\bibitem{mgm_lte_implementation}
N.-D. Nguyen, R.~Knopp, N.~Nikaein, and C.~Bonnet, ``Implementation and validation of multimedia broadcast multicast service for {LTE/LTE}-advanced in {O}pen{A}ir{I}nterface platform,'' in \emph{Proc. of the 38th Annual IEEE Conf. on Local Computer Networks (Workshops)}, Oct. 2013, pp. 70--76.

\bibitem{mgm_80211}
D.~Dujovne and T.~Turletti, ``Multicast in 802.11 {WLAN}s: an experimental study,'' in \emph{Proc. of the 9th ACM Intl. Symp. on Modeling, Analysis and Simulation of Wireless and Mobile Systems}, 2006, p. 130–138.

\bibitem{ZhiguoDing_TSP_NOMA_MMA}
M.~F. Hanif, Z.~Ding, T.~Ratnarajah, and G.~K. Karagiannidis, ``A {M}inorization-{M}aximization method for optimizing sum rate in the downlink of non-orthogonal multiple access systems,'' \emph{{IEEE} Trans. Signal Process.}, vol.~64, no.~1, pp. 76--88, 2016.

\bibitem{Wang_ICCC2018_v2x}
Z.~Wang, J.~Hu, G.~Liu, and Z.~Ma, ``Optimal power allocations for relay-assisted {NOMA}-based {5G V2X} broadcast/multicast communications,'' in \emph{Proc. of the IEEE/CIC Intl. Conf. on Commun. in China (ICCC)}, 2018, pp. 688--693.

\bibitem{Ivari_ICC2020_UCNOMA}
S.~M. Ivari, M.~Caus, M.~A. Vazquez, M.~R. Soleymani, Y.~R. Shayan, and A.~I. Perez-Neira, ``Power allocation and user clustering in {M}ulticast {NOMA} based satellite communication systems,'' in \emph{Proc. of the IEEE Intl. Conf. on Commun. (ICC)}, 2020, pp. 1--6.

\bibitem{Ihsan_NOMAV2X_2021}
A.~Ihsan, W.~Chen, S.~Zhang, and S.~Xu, ``Energy-efficient {NOMA} multicasting system for beyond {5G} cellular {V2X} communications with imperfect {CSI},'' \emph{{IEEE} Trans. Intell. Transp. Syst.}, vol.~23, no.~8, pp. 10\,721--10\,735, 2022.

\bibitem{Choi_NOMA_Multicast_2015}
J.~Choi, ``Minimum power multicast beamforming with superposition coding for multiresolution broadcast and application to {NOMA} systems,'' \emph{{IEEE} Trans. Commun.}, vol.~63, no.~3, pp. 791--800, 2015.

\bibitem{Yi_WCNC2017_Multicast_geocast}
Y.~Zhang, T.-X. Zheng, Q.~Yang, H.-M. Wang, B.~Wang, and Z.~Li, ``The application of {N}on-{O}rthogonal {M}ultiple {A}ccess in {5G} {P}hysical-{L}ayer {M}ulti-{R}egion geocast,'' in \emph{2017 IEEE Wireless Communications and Networking Conference (WCNC)}, 2017, pp. 1--6.

\bibitem{NOMAineffcient}
B.~Clerckx, Y.~Mao, R.~Schober, E.~A. Jorswieck, D.~J. Love, J.~Yuan, L.~Hanzo, G.~Y. Li, E.~G. Larsson, and G.~Caire, ``Is {NOMA} efficient in multi-antenna networks? {A} critical look at next generation multiple access techniques,'' \emph{IEEE Open Journal of the Communications Society}, vol.~2, pp. 1310--1343, 2021.

\bibitem{HanKobayashi}
T.~Han and K.~Kobayashi, ``A new achievable rate region for the interference channel,'' \emph{{IEEE} Trans. Inf. Theory}, vol.~27, no.~1, pp. 49--60, 1981.

\bibitem{mao2022fundmental}
Y.~Mao, O.~Dizdar, B.~Clerckx, R.~Schober, P.~Popovski, and H.~V. Poor, ``{R}ate-{S}plitting {M}ultiple {A}ccess: {F}undamentals, {S}urvey, and {F}uture {R}esearch {T}rends,'' \emph{{IEEE} Commun. Surveys Tuts.}, 2022.

\bibitem{RSMA_JSAC_Primer}
B.~Clerckx, Y.~Mao, E.~A. Jorswieck, J.~Yuan, D.~J. Love, E.~Erkip, and D.~Niyato, ``A {P}rimer on {R}ate-{S}plitting {M}ultiple {A}ccess: {T}utorial, {M}yths, and {F}requently {A}sked {Q}uestions,'' \emph{{IEEE} J. Sel. Areas Commun.}, vol.~41, no.~5, pp. 1265--1308, May 2023.

\bibitem{Mao2018}
Y.~Mao, B.~Clerckx, and V.~O. Li, ``Rate-{S}plitting {M}ultiple {A}ccess for {D}ownlink {C}ommunication {S}ystems: {B}ridging, {G}eneralizing, and {O}utperforming {SDMA} and {NOMA},'' \emph{EURASIP Journal on Wireless Communications and Networking}, vol. 2018, no.~1, p. 133, May 2018.

\bibitem{RSMAUnifying}
B.~Clerckx, Y.~Mao, R.~Schober, and H.~V. Poor, ``{R}ate-{S}plitting {U}nifying {SDMA}, {OMA}, {NOMA}, and {M}ulticasting in {MISO} {B}roadcast {C}hannel: {A} {S}imple {T}wo-{U}ser {R}ate {A}nalysis,'' \emph{{IEEE} Wireless Commun. Lett.}, vol.~9, no.~3, pp. 349--353, 2020.

\bibitem{lyu2023prototype}
X.~Lyu, S.~Aditya, J.~Kim, and B.~Clerckx, ``{R}ate-{S}plitting {M}ultiple {A}ccess: The {F}irst {P}rototype and {E}xperimental {V}alidation of its {S}uperiority over {SDMA} and {NOMA},'' \emph{{IEEE} Trans. Wireless Commun.}, 2024, (early access).

\bibitem{RSMA_overloaded_Onur}
O.~Dizdar, A.~Sattarzadeh, Y.~X. Yap, and S.~Wang, ``{RSMA} for overloaded {MIMO} networks: Low-complexity design for max–min fairness,'' \emph{{IEEE} Trans. Wireless Commun.}, vol.~23, no.~6, pp. 6156--6173, Jun. 2024.

\bibitem{RSMAOverloaded2}
Y.~Mao, E.~Piovano, and B.~Clerckx, ``Rate-splitting multiple access for overloaded cellular internet of things,'' \emph{{IEEE} Trans. Commun.}, vol.~69, no.~7, pp. 4504--4519, 2021.

\bibitem{junliang_RSMAovl_JWCML}
J.~Yang, S.~Gao, Y.~Lu, Z.~Yang, and G.~Tu, ``Joint inter-group common stream superposition and user grouping-based {RSMA} in overloaded systems,'' \emph{{IEEE} Wireless Commun. Lett.}, pp. 1--1, 2024.

\bibitem{Yalcin_2020_JVT}
A.~Z. Yalcin, M.~Yuksel, and B.~Clerckx, ``Rate splitting for multi-group multicasting with a common message,'' \emph{{IEEE} Trans. Veh. Technol.}, vol.~69, no.~10, pp. 12\,281--12\,285, Oct 2020.

\bibitem{Yin_2021_TCOM}
L.~Yin and B.~Clerckx, ``Rate-splitting multiple access for multigroup multicast and multibeam satellite systems,'' \emph{{IEEE} Trans. Commun.}, vol.~69, no.~2, pp. 976--990, Feb 2021.

\bibitem{Hongzhi_JBC}
H.~Chen, D.~Mi, T.~Wang, Z.~Chu, Y.~Xu, D.~He, and P.~Xiao, ``Rate-splitting for multicarrier multigroup multicast: Precoder design and error performance,'' \emph{{IEEE} Trans. Broadcast.}, vol.~67, no.~3, pp. 619--630, 2021.

\bibitem{hongzhi_RSMAovl_icc}
H.~Chen, D.~Mi, Z.~Liu, P.~Xiao, and R.~Tafazolli, ``Rate-{S}plitting for overloaded multigroup multicast: Error performance evaluation,'' in \emph{Proc. of the {IEEE} Intl. Conf. on Comms. {(ICC)} Workshops}, 2020, pp. 1--6.

\bibitem{Yin_ICC_2021}
L.~Yin, O.~Dizdar, and B.~Clerckx, ``Rate-splitting multiple access for multigroup multicast cellular and satellite communications: {PHY} layer design and link-level simulations,'' in \emph{Proc. of the {IEEE} {I}ntl. {C}onf. on Communications ({ICC}) Workshops}, 2021, pp. 1--6.

\bibitem{Cui_etal_2023_COML}
H.~Cui, L.~Zhu, Z.~Xiao, B.~Clerckx, and R.~Zhang, ``Energy-efficient rsma for multigroup multicast and multibeam satellite communications,'' \emph{{IEEE} Wireless Commun. Lett.}, vol.~12, no.~5, pp. 838--842, May 2023.

\bibitem{HamdiMISOImperfectCSIT}
H.~Joudeh and B.~Clerckx, ``Sum-rate maximization for linearly precoded downlink multiuser {MISO} systems with partial {CSIT}: {A} rate-splitting approach,'' \emph{{IEEE} Trans. Commun.}, vol.~64, no.~11, pp. 4847--4861, 2016.

\bibitem{Christensen_etal_2008}
S.~S. Christensen, R.~Agarwal, E.~De~Carvalho, and J.~M. Cioffi, ``Weighted sum-rate maximization using weighted mmse for mimo-bc beamforming design,'' \emph{{IEEE} Trans. Wireless Commun.}, vol.~7, no.~12, pp. 4792--4799, Dec. 2008.

\bibitem{mosquera2023linkadaption}
C.~Mosquera and F.~G{\'o}mez-Cuba, ``Link {A}daptation for {R}ate {S}plitting {S}ystems with partial {CSIT},'' \emph{{IEEE} J. Sel. Areas Commun.}, vol.~41, no.~5, pp. 1336--1350, May 2023.

\bibitem{FinephaseShifting}
H.~Minn, ``A robust timing and frequency synchronization for {OFDM} systems,'' \emph{{IEEE} Trans. Wireless Commun.}, vol.~2, no.~4, pp. 822--839, 2003.

\bibitem{trifonovPolar}
P.~Trifonov, ``{E}fficient {D}esign and {D}ecoding of {P}olar {C}odes,'' \emph{{IEEE} Trans. Commun.}, vol.~60, no.~11, pp. 3221--3227, 2012.

\bibitem{constructionPolar}
H.~Li and J.~Yuan, ``{A} {P}ractical {C}onstruction {M}ethod for {P}olar {C}odes in {AWGN} {C}hannels,'' in \emph{IEEE 2013 Tencon - Spring}, 2013, pp. 223--226.

\bibitem{listdecoding}
I.~Tal and A.~Vardy, ``{L}ist {D}ecoding of {P}olar {C}odes,'' \emph{{IEEE} Trans. Inf. Theory}, vol.~61, no.~5, pp. 2213--2226, 2015.

\end{thebibliography}
